\documentclass[12pt]{article}
\usepackage[margin=1in]{geometry}
\usepackage{graphicx}
\usepackage{graphbox}

\usepackage{amssymb}
\usepackage{amsmath}
\usepackage{amsfonts}
\usepackage{float}
\usepackage{natbib}
\usepackage{xr}
\usepackage{hyperref}
\usepackage{authblk}
\usepackage{cleveref}

\newcommand{\btheta}{ \mbox{\boldmath $ \theta $} }

\newcommand{\bs}{\textbf{s}}
\newcommand{\bY}{\textbf{Y}}

\begin{document}

\title{Spatial modeling of day-within-year temperature time series: an examination of daily maximum temperatures in Arag{\'o}n, Spain}

\author[1,*]{Jorge Castillo-Mateo} 
\author[1]{Miguel Lafuente}
\author[1]{Jes{\'u}s As{\'i}n}
\author[1]{Ana C. Cebri{\'a}n}
\author[2]{Alan E. Gelfand}
\author[1]{Jes{\'u}s Abaurrea}
\affil[1]{Department of Statistical Methods, University of Zaragoza, Zaragoza, Spain}
\affil[2]{Department of Statistical Science, Duke University, Durham, North Carolina, USA}
\affil[*]{Correspondence: jorgecm@unizar.es}

\date{}

\maketitle

\begin{abstract}

Acknowledging a considerable literature on modeling daily temperature data, we propose a multi-level spatio-temporal model which introduces several innovations in order to explain the daily maximum temperature in the summer period over $60$ years in a region containing Arag{\'o}n, Spain. The model operates over continuous space but adopts two discrete temporal scales, year and day within year.  It captures temporal dependence through autoregression on days within year and also on years. Spatial dependence is captured through spatial process modeling of intercepts, slope coefficients, variances, and autocorrelations.  The model is expressed in a form which separates fixed effects from random effects and also separates space, years, and days for each type of effect. 
Motivated by exploratory data analysis, fixed effects to  capture the influence of elevation, seasonality  and  a  linear trend are employed. Pure errors are  introduced for years, for locations within years, and for locations at days within years. The  performance of the model is checked using a leave-one-out cross-validation. Applications of the  model are presented including prediction of the daily temperature series at unobserved  or partially observed sites and inference to investigate climate change comparison.


\end{abstract}

\section{Introduction}

Evidence of global warming in the climate system is strong and many of the observed changes since the 1950s are unprecedented, with an estimated anthropogenic increase of 0.2$^{\circ}$C per decade due to past and ongoing emissions \citep{IPCC2013,IPCC2018}.  Climate change raises significant  concerns as it may result in health problems and death,  degradation of flora and fauna biodiversity,  reductions in crop production and   increase of pests, etc. In this framework, the analysis of daily maximum temperatures and their long-term trends  over time is particularly important due to the strong potential impact on public health \citep{roldan2016effect,rossati2017,watts2015}, agriculture \citep{hatfield2011,schlenker2009}, and  economy \citep{diffenbaugh2019}.

We propose a new multi-level spatio-temporal model to explain the  daily maximum temperature  in the summer period, in an area containing  the Comunidad Aut{\'o}noma de Arag{\'o}n in the northeast of Spain. 
The region includes part of the Ebro Valley in the center, with mountainous areas in the south (Iberian System) and north (Pyrenees). The valley is an extensively irrigated production  area with garden crops, fruits and vegetables, as well as rainfed agriculture with cereals, almonds, wine and oil. In the mountainous areas there are  some protected natural spaces  with extensive forests and a high diversity of landscapes. It is an area of great biodiversity  with important water resources for the region. Despite its relatively small size, spatio-temporal modeling of  the temperatures in this region is a challenge  due to the heterogeneous orography  and the climatic variability.

The spatio-temporal model seeks to characterize spatial patterns and  detect trends over time in the daily maximum  temperature during the summer period. It is specified over continuous space but adopts two discrete units of time, years and days within years. This allows us to model  the  time evolution  of daily maximum temperatures during the summer, omitting the cooler months that are not of interest here.  The model introduces temporal dependence using autoregression terms for days within years and also for years. The model separates fixed and random effects in the mean.  Fixed effects capture the global mean, the seasonal component across days, the average long-term trend across years, and the influence of elevation. Random effects are employed for the spatial dependence in the intercepts, the slope coefficients, the autoregression coefficients, and the variances of the responses. The two temporal scales allow us to separate space, years, and days within years for each type of effect.  Three pure error processes are adopted, one for locations at days within years, one for locations within years, and one for years. The full specification is motivated by exploratory analyses.  Altogether, the model provides a better understanding of the temporal evolution of temperatures for the entirety of the region along with the spatial uncertainty  linked to those features.

The model is specified in a hierarchical Bayesian framework and estimated using a Markov chain Monte Carlo (MCMC) algorithm. In this framework, posterior predictive distributions for the features of daily maximum temperatures (trends, persistence, mean, variance, etc.) can be readily obtained. In particular, we can obtain posterior predictive samples of the spatial processes and  the daily  maximum temperature series at unobserved sites. Prediction at unobserved sites is particularly important in Arag{\'o}n since this region is sparsely monitored due to rural depopulation; there is a lack of observed series in many areas of interest.  The model can also be used to impute periods of missing observations in a series.

Space-time modeling of environmental series has received substantial attention in the literature. \cite{Sahu06} proposed a random effects model for fine particulate matter concentrations in the midwestern United States. \cite{sahu2007} proposed a space-time hierarchical model for daily 8-hour maximum ozone levels in the state of Ohio. This model includes an autoregressive part for the residuals of the fixed effects, a global annual intercept and a spatially correlated error term. \cite{lemos2007} modeled monthly water temperature data in a Central California Estuary. They used a Bayesian approach to separate the seasonal cycle, short-term fluctuations, and long-term trends by means of local mixtures of two patterns.   With regard to temperature  models, \cite{Craigmile11} built space‐time hierarchical Bayesian models using daily mean temperatures in Central Sweden that emphasize modeling trend through a wavelet specification, as well as seasonality, and error that may exhibit space‐time long‐range dependence.  \cite{verdin2015coupled}  modeled maximum and minimum temperature to develop a weather generator  using spatial Gaussian processes (GPs),  where both temperature models are autoregressive with spatially varying model coefficients and spatial correlation.  \cite{Sha20} proposed a three step space-time regression-kriging model for monthly  average temperature data.  With such data, they first remove seasonality, then they regress the revised data on environmental predictors, and finally they take the resulting residuals and administer spatio-temporal variogram modeling. By contrast,  models for daily temperatures take a different approach, seeking to explicitly express short-term persistence of temperature.  They employ autoregressive terms, e.g., the one-point model   by  \cite{mohammadi2020developing}. A modeling approach very different from our mean specification considers extremes in the daily temperature series and leads to extreme value modeling under the block maxima framework or peaks-over-threshold framework \citep[see, e.g.,][]{reich2014,bopp2017}.

The outline of the paper is as follows. An exploratory analysis  to motivate the complexity of the model is given in Section~\ref{sec:data}. Section~\ref{sec:model} describes the modeling details, and Section~\ref{sec:analysis} presents a leave-one-out cross-validation (LOOCV) analysis for model comparison as well as some results and applications for the selected model.  Section~\ref{sec:summary} ends the paper with some conclusions and future work. Supplementary Materials accompanying this paper appear online.

\section{Data and exploratory analysis}
\label{sec:data}

\begin{figure}
\begin{center}
\includegraphics[align=c,width=.56\textwidth]{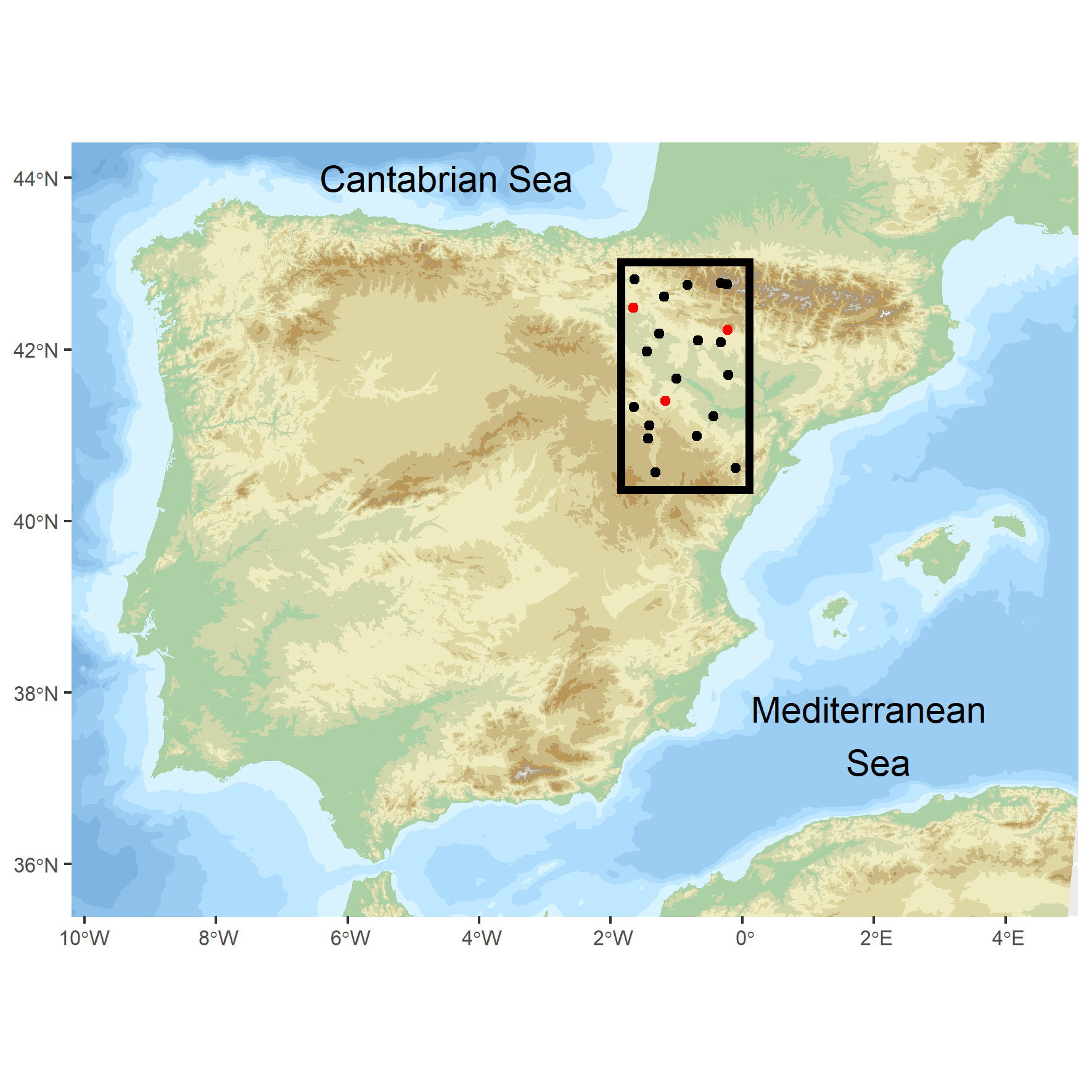}
\includegraphics[align=c,width=.39\textwidth]{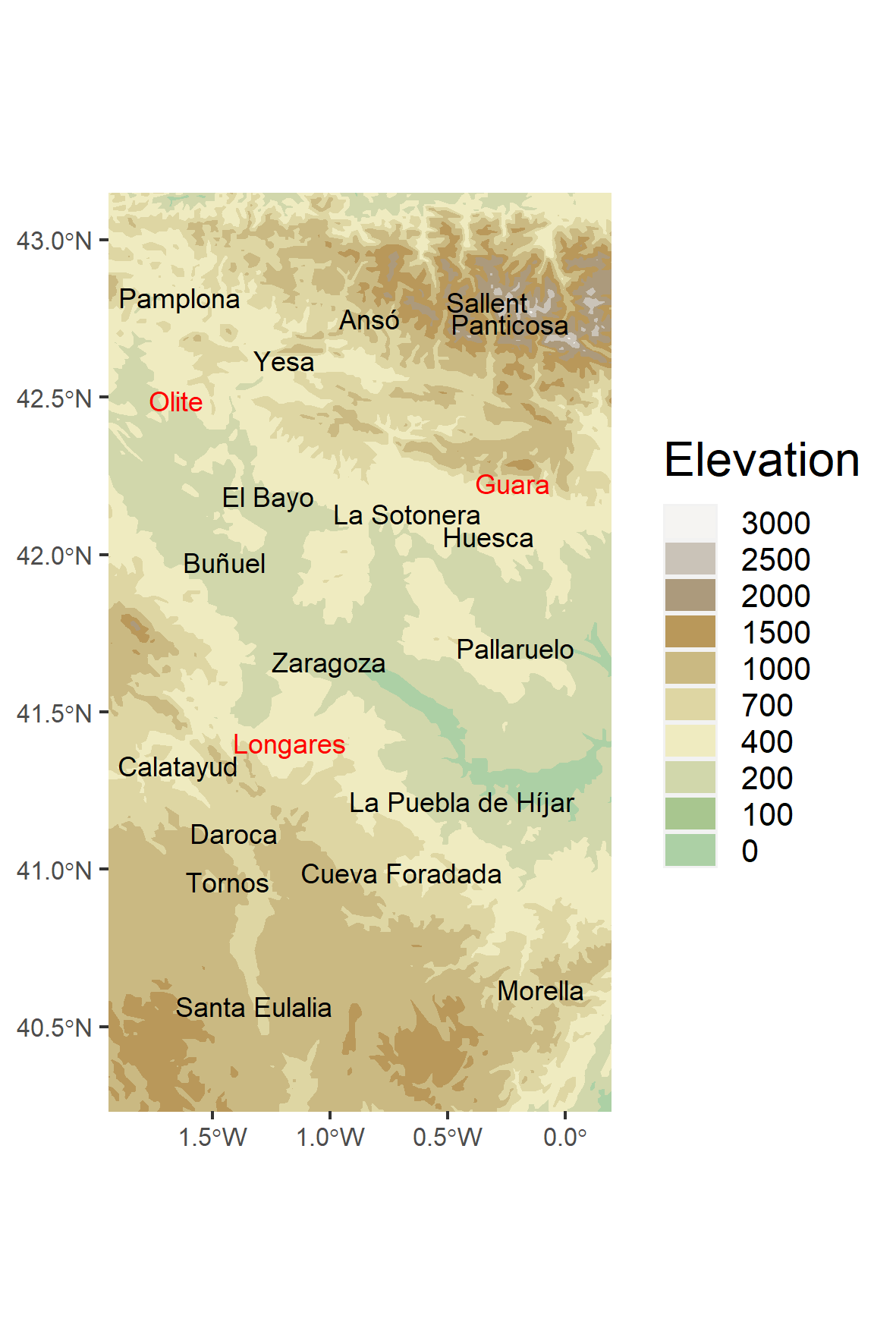}
\vspace*{-10mm}\caption{Map locating within the Iberian Peninsula the 18 sites (black) used to fit the model and the 3 unobserved sites (red) where prediction is carried out. \label{fig:Map}}
\end{center}
\end{figure}

The point-referenced dataset we use contains 18 daily maximum temperature observational series from AEMET (the Spanish Meteorological Office) around the Comunidad Aut{\'o}noma de Arag{\'o}n (see Figure~\ref{fig:Map}).  The time series include the daily observations from May to September (MJJAS), corresponding to the extended summer period, and span  the period from 1956 to 2015.  The region of interest is located in the central portion of the Ebro Basin in the northeastern part of Spain and has an area of 53,279 km$^2$, wherein the areas above 500 m and 1,000 m are 32,924 km$^2$ and 15,195 km$^2$ respectively.  The maximum elevation is roughly 3,400 m in the Pyrenees, 2,600 m in the Iberian System, and between 200 and 400 m in the Central Valley. Most of the area is characterized by a Mediterranean-Continental dry climate with irregular rainfall and a large temperature range. However,  climate differences can be distinguished by elevation and the  influence from the Mediterranean Sea in the east as well as the continental conditions of the Iberian Central Plateau in the southwest \citep{AEMET11}.

\begin{figure}
\begin{center}
\includegraphics[width=\textwidth]{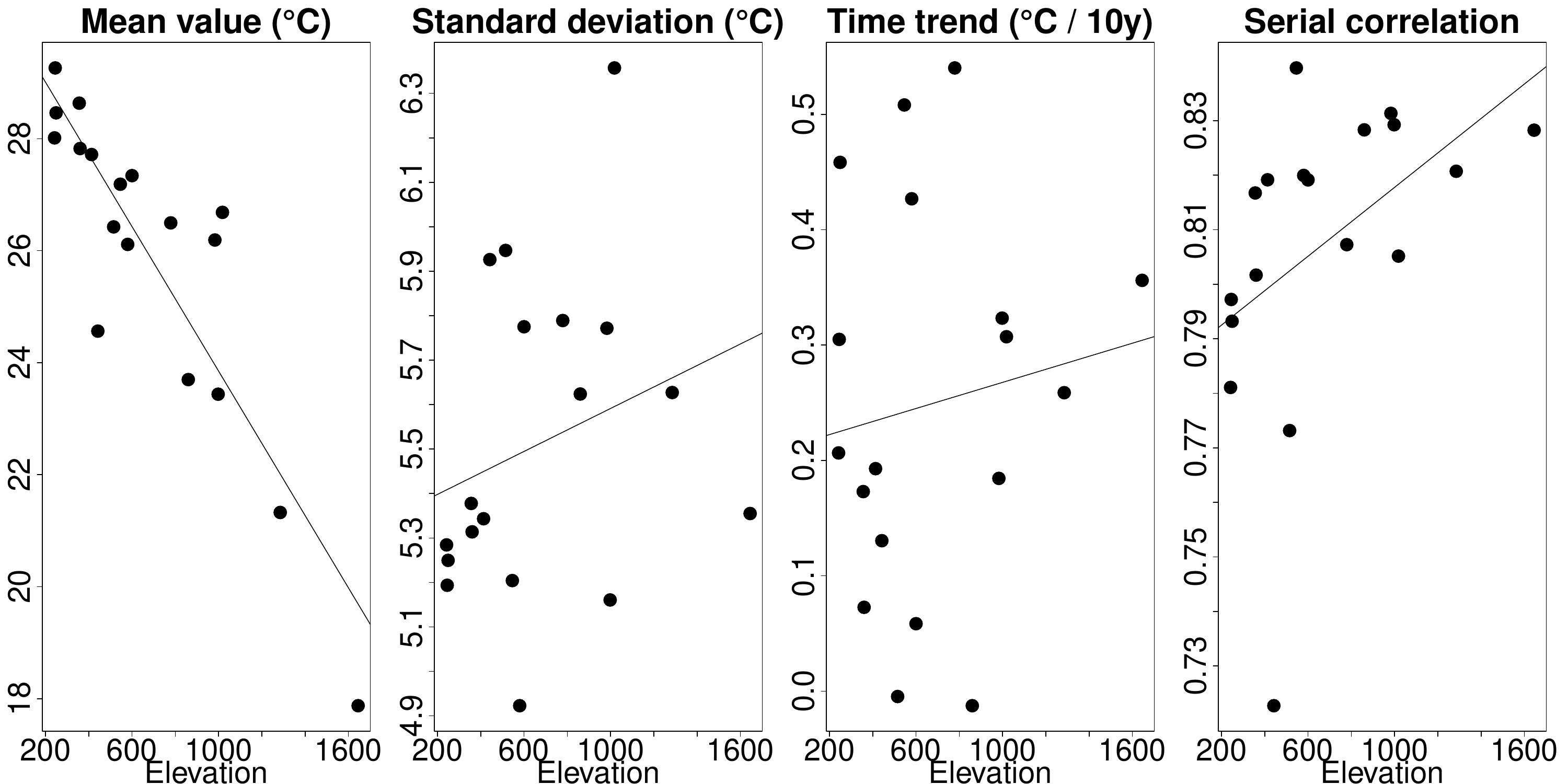} \\ \vspace{5mm}
\includegraphics[width=\textwidth]{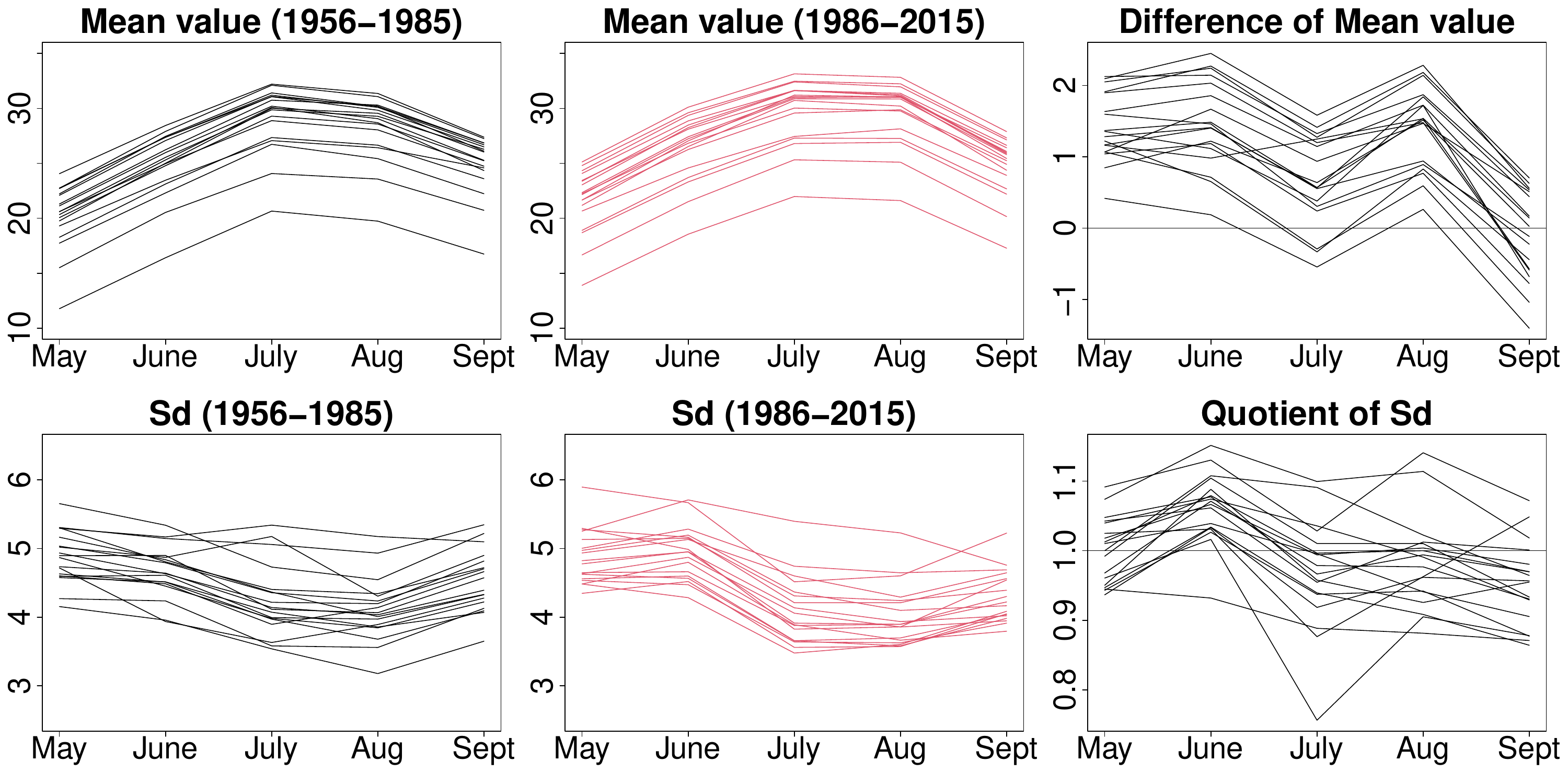}
\caption{Top: Mean value, standard deviation, annual time trend, and serial correlation against elevation for the daily maximum temperature series at the 18 sites. Bottom: Mean value and standard deviation of the series in both 30-year periods, 1956--1985 and 1986--2015, and  the change between them, expressed as differences for the mean and quotients for the standard deviation. \label{fig:mean:altitude:change:mean:stdev}}
\end{center}
\end{figure}

We summarize an extensive exploratory data analysis of the daily maximum temperature series that helps us establish the covariates and spatio-temporal structures that are candidates for inclusion in the model. The top plots in Figure~\ref{fig:mean:altitude:change:mean:stdev} show the variability in temperature characteristics and the influence of elevation on them. The two plots on the left show the mean and the standard deviation of temperature at each site against elevation.  The mean temperature  shows an approximately linear decreasing relation with  elevation, varying from  almost $30$ to $18^{\circ}$C. However, there exist other influential  factors, e.g., Sallent in the north and Tornos in the south have  both an elevation around  1,000 m but a quite lower mean temperature is observed for the latter (see Table~S1 in Supplementary Materials).

The bottom plots in Figure~\ref{fig:mean:altitude:change:mean:stdev} summarize the mean and standard deviation from data corresponding to a month in MJJAS for the 18 sites in the periods 1956--1985 and 1986--2015; the summary measures are calculated in 30-year periods following the recommendation of the \cite{WMO17}. The seasonal pattern for  all of the series is quite similar, i.e., the maximum mean temperature is observed in July and the minimum in May, with a difference of around $7^{\circ}$C between them. The  range of the mean temperatures among sites is around $10^{\circ}$C, so the spatial variability of the mean is a bit higher than the variability  at each site within the summer. The mean of the set of standard deviations is slightly higher than $4^{\circ}$C.  However, relevant spatial differences are observed  with a range of values  around $1.5^{\circ}$C.  Temporal variability is lower within the summer.

To explore the effect of global warming in the region, the changes between 1956--1985 and 1986--2015 periods, expressed as differences for the means and quotients for the standard deviations, are also shown on the bottom-right plot in 
Figure \ref{fig:mean:altitude:change:mean:stdev} and Table~S1.  The  mean temperature  in 1986--2015  has increased from 1956--1985 by roughly  $1^{\circ}$C, with a slightly smaller increase in  the northeastern sites.  The increase of the mean temperature is observed in May, June, August and, except for three sites, in July. No relevant change in the seasonal pattern is observed. The spatial variability  in the two periods is  similar. As for the standard deviations, no evidence of temporal change is observed, with all of the quotients between the two periods being approximately one.

The two plots on the top-right in Figure~\ref{fig:mean:altitude:change:mean:stdev} summarize an exploratory analysis of the behavior of the time series over time. The first shows the slope regressed against year (expressed in $^{\circ}$C per decade), fitted by ordinary least squares to the daily maximum temperature series in each site. Clear differences are observed in the 18 fitted trends, suggesting the need to include a spatial random effect to reflect this feature.  The variability in the trends does not seem to be related to the elevation. The last plot shows the serial correlation in the temperature series. A strong correlation, higher than 0.72, is observed for all the sites but with spatial differences. The strong autocorrelation is probably caused by a persistent anticyclonic situation that tends to affect the Iberian Peninsula in the summer. Sites with a higher elevation seem to show a slightly higher persistence.

As an additional exploratory analysis, 18 hierarchical temporal models were fitted, one for each of the available  sites.  These local models, which are summarised in Section~S1.1 of the Supplementary Materials, are useful to  identify  the time structures required for the temperature series  and to evaluate  the spatial variability of  the fitted terms. The results motivate the introduction of spatially varying intercepts, trends, autoregression coefficients, and variances for the spatial variability in the model.

\section{The Model} \label{sec:model}

We propose a multi-level (i.e., hierarchical) full mean model for daily maximum temperatures that operates over continuous space and two discrete temporal scales. 
It captures temporal dependence through autoregression on days within year and on years.  It captures spatial dependence through spatial process modeling of intercepts, slope coefficients, variances, and autocorrelations. 
We detail this model below and then discuss model fitting, prediction under the model, and model comparison.

\subsection{Model construction}

Let $Y_{t\ell}(\bs)$ denote the daily maximum temperature for day $\ell$, $\ell=2,\ldots,L$ of year $t$, $t=1,\ldots,T$ at location $\bs \in D$, where $D$ is our study region.  Here, for all years, $\ell=1$ corresponds to May 1 and $L=153$ corresponds to September 30.  It is convenient to express the full model in a form which separates fixed effects from random effects and also carefully separates space, years, and days for each type of effect.  Specifically, we model daily maximum temperature for day $\ell$, year $t$, and location $\bs$ by
\begin{equation} \label{eq:model}
Y_{t\ell}(\bs)= \mu_{t\ell}(\bs; \btheta_{f}) +  \gamma_{t}(\bs) + \rho_{Y}(\bs)\left(Y_{t,\ell-1}(\bs) - (\mu_{t,\ell-1}(\bs; \btheta_{f}) + \gamma_{t}(\bs))\right) + \epsilon_{t\ell}^{(Y)}(\bs).
\end{equation}

Here, $ \mu_{t\ell}(\bs; \btheta_{f})$ denotes the fixed effects component and $\gamma_{t}(\bs)$ the random effects component.  We specify
\begin{equation} \label{eq:fixed}
\mu_{t\ell}(\bs; \btheta_{f}) = \beta_0 + \alpha t + \beta_1 \text{sin}(2\pi \ell/365) +  \beta_2 \text{cos}(2\pi \ell/365) + \beta_3 \text{elev}(\bs)
\end{equation}
in which $\beta_{0}$ is a global intercept, $\alpha$ is a global linear trend coefficient, the $\sin$ and
$\cos$ terms are introduced to provide an annual seasonal component, and $\text{elev}(\bs)$ is the elevation at $\bs$. We denote these \emph{fixed} effect parameters by $\btheta_{f} = (\beta_{0}, \alpha, \beta_1, \beta_2, \beta_3)$.

We specify
\begin{equation} \label{eq:random}
\gamma_{t}(\bs) = \beta_{0}(\bs) + \alpha(\bs)t + \psi_{t} + \eta_{t}(\bs).
\end{equation}

In \eqref{eq:random}, $\psi_{t}$ follows an AR(1) specification, i.e., $\psi_{t} = \rho_{\psi}\psi_{t-1} + \lambda_{t}$, providing an autoregression in years for annual intercepts.  This autoregression could help to capture factors yielding correlation across years such as the influence of variation in solar activity on the earth's surface temperature or the El Ni{\~n}o Southern Oscillation.  However, in Section~\ref{sec:fitting}, we discover that $\rho_{\psi}$ is not significantly different from $0$. We still need the $\psi$'s in the model to address the fact that some years are  warmer or colder than others but we do not need to specify them autoregressively.  We denote the variance for this component by $\sigma^{2}_{\lambda}$.

Continuing, $\beta_{0}(\bs)$ is a mean-zero GP with an exponential covariance function having variance parameter $\sigma^{2}_{\beta_{0}}$ and decay parameter $\phi_{\beta_{0}}$, and $\alpha(\bs)$ is a mean-zero GP with an exponential covariance function having variance parameter $\sigma^{2}_{\alpha}$ and decay parameter $\phi_{\alpha}$.  Thus, $\beta_{0}(\bs)$ provides local spatial adjustment to the intercept and $\alpha(\bs)$ provides local slope adjustment to the linear trend.  Due to the simplicity of linear time trends they are often used in climate studies  \citep{IPCC2013}.  Here, they provide an extremely flexible, locally linear baseline specification. Further, we add local space-time varying random effects, $\eta_t(\bs)$, to provide adjustment to this baseline. We collect the random effects parameters into $\btheta_{r}= (\rho_{\psi}, \sigma^{2}_{\lambda}, \sigma^{2}_{\beta_{0}}, \phi_{\beta_{0}}, \sigma^{2}_{\alpha}, \phi_{\alpha})$.

The entire specification is supplied distributionally in the form of a multi-level hierarchical model as
\begin{equation} \label{eq:hierarchy}
\begin{gathered}\relax
[Y_{t\ell}(\bs) \mid Y_{t,\ell-1}(\bs), \btheta_{f}, \gamma_{t}(\bs), \rho_{Y}(\bs), \sigma^{2}_{\epsilon}(\bs)]\\
[\gamma_{t}(\bs) \mid \beta_{0}(\bs), \alpha(\bs), \psi_{t},  \sigma^{2}_{\eta}]\\
[\beta_{0}(\bs)|\sigma_{\beta_{0}}^{2}, \phi_{\beta_{0}}]
[\alpha(\bs)|\sigma_{\alpha}^{2}, \phi_{\alpha}]
[\psi_{t}|\psi_{t-1}, \rho_{\psi}, \sigma_{\lambda}^{2}]
[Z_{\rho_{Y}}(\bs)|Z_{\rho_Y},\sigma^{2}_{\rho_{Y}}, \phi_{\rho_{Y}}]
[Z_{\sigma_\epsilon^2}(\bs)|Z_{\sigma^{2}_{\epsilon}}, \sigma^{2}_{\sigma^{2}_{\epsilon}},\phi_{\sigma^{2}_{\epsilon}}]\\
[\btheta_f] [\btheta_r] [\sigma_\eta^2]
[Z_{\rho_Y}] [\sigma_{\rho_Y}^2] [\phi_{\rho_Y}]
[Z_{\sigma_\epsilon^2}] [\sigma_{\sigma_\epsilon^2}^2] [\phi_{\sigma_\epsilon^2}].
\end{gathered}
\end{equation}

As a result, we have introduced three pure error terms: $\lambda_{t} \stackrel{iid}{\sim} N(0, \sigma_{\lambda}^{2})$ at yearly scale, $\eta_{t}(\bs) \stackrel{iid}{\sim} N(0, \sigma_{\eta}^{2})$ at sites within years, and $\epsilon^{(Y)}_{t\ell}(\bs) \stackrel{ind.}{\sim} N(0, \sigma_{\epsilon}^{2}(\bs))$ at sites for days within years.
Additionally, $\rho_Y(\bs)$ and $\sigma_{\epsilon}^2(\bs)$ are, respectively, a spatially varying autoregressive term and a spatially varying variance at location $\bs$, both of which are assumed constant over days and years.  We model $\log\left\{(1+\rho_Y(\bs))/(1-\rho_Y(\bs))\right\} = Z_{\rho_Y}(\bs) \sim GP(Z_{\rho_Y},  C(\cdot;\sigma_{\rho_Y}^2,\phi_{\rho_Y}))$, and $\log\{\sigma_\epsilon^2(\bs)\} = Z_{\sigma_\epsilon^2}(\bs) \sim GP(Z_{\sigma_\epsilon^2}, C(\cdot;\sigma_{\sigma_{\epsilon}^2}^2,\phi_{\sigma_{\epsilon}^2}))$, again with exponential covariance functions. Motivation for adopting spatially varying specifications for these terms arises from exploratory data analysis at the level of the individual sites.   That is, suppose we fit the model above but ignore spatial structure and treat the sites as conditionally independent. We show in Section~S1.1 of the Supplementary Materials that the assumptions of constant autoregression coefficients and constant variances over the region do not seem justified.

\begin{figure}[t]
	\centering
	\includegraphics[width=0.8\textwidth, trim=4.5cm 9.5cm 3cm 4.5cm, clip=TRUE]{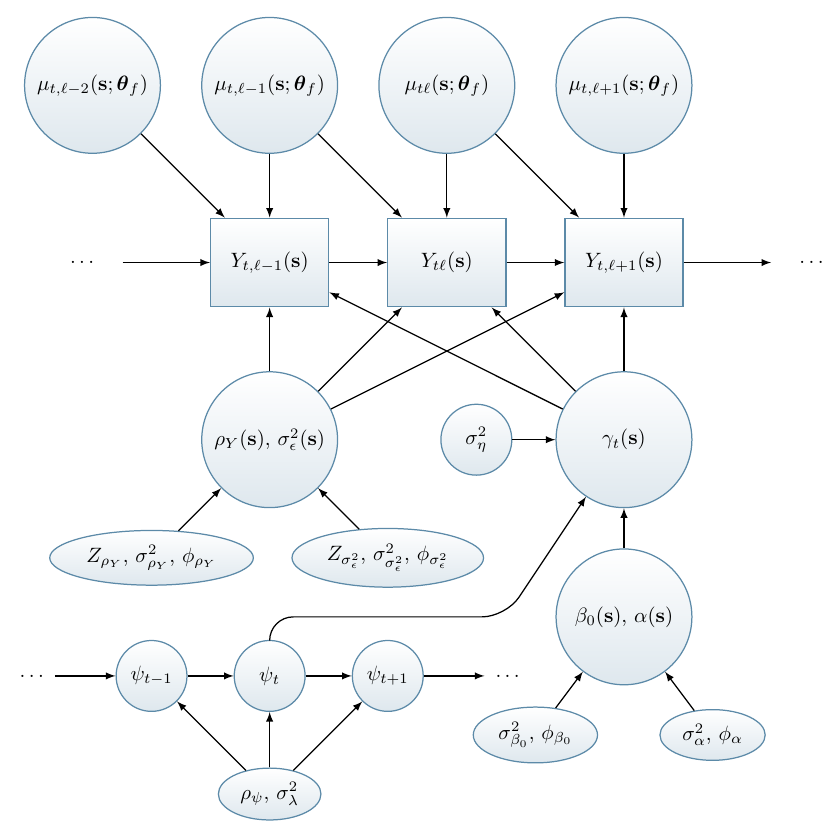}
	\caption{Graphical model for specification in \Cref{eq:model,eq:fixed,eq:random,eq:hierarchy}. Rectangular nodes are observed, circular nodes are unobserved. \label{graph}}
\end{figure}

All of the components considered in the full model and their relationships are depicted in the graphical model in Figure~\ref{graph}.  This diagram, perhaps, reveals the complexity of the full model more readily than through \Cref{eq:model,eq:fixed,eq:random,eq:hierarchy}.

The reader might wonder if the GPs above are independent.  We investigated dependence between the intercept and slope GPs using the following coregionalization \citep[][Chapter 9]{BCG}. Suppose $v_1(\textbf{s})$ and $v_2(\textbf{s})$ are independent GPs with zero mean and unit variance whose exponential covariance functions have decay parameters $\phi_1$ and $\phi_2$, respectively. In the full model we insert $\beta_0(\textbf{s}) = a_{11} v_1(\textbf{s})$ and $\alpha(\textbf{s}) = a_{21} v_1(\textbf{s}) + a_{22} v_2(\textbf{s})$. Here, we let $a_{11}$ and $a_{22}$ each have a half (or folded) Gaussian prior while $a_{21}$ has a regular Gaussian prior. The parameter $a_{21}$ captures the dependence between the two processes.  That is, the induced covariance between $\beta_0(\textbf{s})$ and $\alpha(\textbf{s})$ is $a_{21}a_{11}$.  We care whether $a_{21}$ is significantly different from zero with little interest in exactly what the correlation is. Under the model above, the posterior distribution of $a_{21}$ was centered at zero with wide credible intervals.
So, this dependence was not included in the final model for which we present the inference.

Returning to the full model, notice that we have separated the fixed effects according subscripts $t$, $\ell$, and $\bs$.  As for $\gamma_{t}(\bs)$, we can see that it has a spatially varying intercept, a spatially varying coefficient for drift, and an AR(1) model for years.  Also, $\gamma_{t}(\bs)$ has both space and time dependence and,  in fact, we can readily calculate $\text{cov}(\gamma_{t}(\bs), \gamma_{t+h}(\bs'))$.  Under independence of the intercept and slope processes, the \emph{equilibrium} covariance becomes 
\begin{equation}
    \text{cov}(\gamma_{t}(\bs), \gamma_{t+h}(\bs')) =
    C(||\bs-\bs'||; \sigma_{\beta_{0}}^{2},\phi_{\beta_{0}}) + t(t+h) C(||\bs - \bs'||; \sigma_{\alpha}^{2},\phi_{\alpha}) + \frac{\sigma_{\lambda}^{2}}{1- \rho_{\psi}^{2}} \rho_{\psi}^{|h|}.
\end{equation}

Finally, special cases of interest include: $\beta_{0}(\bs)=0$ implies a constant intercept over space, $\alpha(\bs)=0$ implies a constant linear drift over space, $\rho_{\psi}=0$ implies no yearly autoregression.  These assumptions merely revise the form of $\gamma_{t}(\bs)$.  We might consider conditioning on a longer history of maximum temperatures. We experimented with introducing additional lags in the modeling but we found no gain in predictive performance.  We could also consider additional fixed effects, e.g., longitude, latitude or distance to coast, or even adding interactions, e.g., $t \times \text{elev}(\bs)$.  However, the exploratory analysis did not reveal a relationship between daily temperatures and these fixed effects, so they were not introduced in the full model.

\subsection{Model fitting}
\label{sec:fitting}

Model inference is implemented in a Bayesian framework, requiring prior distributions for each of the model parameters. In general, diffuse and, when available, conjugate prior distributions are chosen. Recall that the model adopts a conditional Gaussian distribution for all $Y_{t\ell}(\bs)$'s. Thus it is appropriate to assign each of the coefficient parameters $\beta_0$, $\alpha$, $\beta_1$, $\beta_2$ and $\beta_3$, independent and diffuse Gaussian prior distributions with mean $0$ and standard
deviation $100$.  The variance parameters, $\sigma_\lambda^2$ and $\sigma_\eta^2$, are assigned independent Inverse-Gamma$(2,1)$ prior distributions. In preliminary analyses, the autoregresive term between years, $\rho_\psi$, was assigned a non-informative Uniform$(-1,1)$ prior distribution. As its posterior distribution was centered at zero with wide credible intervals, we set the parameter at $\rho_\psi = 0$. For identifiability, the random effect for the first year, $\psi_1$, is fixed to zero.

Hyperpriors are assigned to the mean of both $Z_{\rho_Y}(\bs)$ and $Z_{\sigma_\epsilon^2}(\bs)$.  That is,  $Z_{\rho_Y}$ and $Z_{\sigma_\epsilon^2}$ are given a Gaussian prior distribution with mean $0$ and standard deviation $100$ and $1$, respectively. The variance parameter for each of the four spatial covariance functions, $\sigma_{\beta_0}^2$, $\sigma_{\alpha}^2$, $\sigma_{\rho_Y}^2$ and $\sigma_{\sigma_\epsilon^2}^2$, is assigned an independent Inverse-Gamma$(2,1)$ prior distribution. Preliminary analyses with a discrete uniform prior distribution for each of the spatial decay parameters indicated that these parameters almost always placed most mass on the smallest decay value. Due to the fact that, with an exponential covariance function, the variance and the decay parameter cannot be individually identified \citep{zhang2004}, and the decay parameter is $3/\text{range}$, we set $\phi \equiv \phi_{\beta_0} = \phi_{\alpha} = \phi_{\rho_Y} = \phi_{\sigma_\epsilon^2} = 3 / d_{max}$, where $d_{max}$ is the maximum distance between any pair of spatial locations.

MCMC is used to obtain samples from the joint posterior distribution. The sampling algorithm is a Metropolis-within-Gibbs version. Since we only have $18$ sites, we fit the model without marginalization over the spatial random effects. Also, we introduce $\tilde{\beta}_0(\bs) = \beta_0 + \beta_0(\bs)$ and $\tilde{\alpha}(\bs) = \alpha + \alpha(\bs)$ within $\gamma_t(\bs)$ for the fitting to enable the benefits of hierarchical centering in the model fitting \citep{gelfand1995}. Details of the MCMC used for the model fitting are provided in Section~S2.1 of the Supplementary Materials. All the covariates have been centered and scaled to have zero mean and standard deviation one to improve the mixing behavior of the algorithm.

\subsection{Spatial and spatio-temporal prediction}

Under the full model, prediction at location $\bs_0$, day $\ell'$, and year $t'$ is based on the posterior predictive distribution of $Y_{t'\ell'}(\bs_0)$ arising from the full model. Here, $\bs_0$ may correspond to a fully observed location (held out for validation), a partially observed location (for completion of a record), or a new location in $D$.  Our goal is not forecasting, so we restrict ourselves to the observed time period $\ell' = 2,\ldots,L$ and $t' = 1,\ldots,T$. Within the Bayesian framework, the posterior predictive distribution for $Y_{t'\ell'}(\bs_0)$ is obtained by integrating over the parameters with respect to the joint posterior distribution.  The formal expression for the posterior predictive distribution for $[Y_{t'\ell'}(\bs_0) \mid \bY]$, where $\bY$ is the observed data, is given in Section~S2.2 of the Supplementary Materials. Customarily, the distribution is obtained empirically through posterior samples. That is, with MCMC algorithms, samples of the posterior parameters are used to obtain posterior predictions of observations, so-called composition sampling \citep[see][Chapter 6; and Section~S2.2 for the details]{BCG}.

\subsection{Model evaluation}
\label{sec:metrics}

For model assessment, a LOOCV is carried out to compare the spatial predictive performance of the models. The full model considered includes four spatial GPs.  To validate that model as well as the importance of the considered GPs,  reduced models incorporating 0, 1, 2, or 3 GPs are fitted. Models are presented explicitly in Section~\ref{sec:metrics_results} where we further clarify that removing particular terms allows explicit interpretation of the resulting reduced models.

Results from Section~\ref{sec:metrics_results} favors the full model and so, results for this model are presented subsequently.  However, several of the reduced models yield essentially equivalent  global performance, though the fit at some sites is  poorer. We attempt to clarify why this might be expected but also show that each set of random effects reveals differences across sites, further encouraging us to retain them in the inference presentation.

For each location in the hold-out set, the entire time series of daily maximum temperatures is withheld during model fitting.  Then, for location $\bs_{i}$, we conduct our model comparison through the following metrics: (i) root mean square error (RMSE),  (ii) mean absolute error (MAE), (iii) continuous ranked probability score \citep[CRPS;][]{gneiting2007}, and (iv) coverage (CVG).  By definition,
\begin{align*}
\text{RMSE}_i &= \sqrt{\frac{1}{T (L-1)}	\sum_{t=1}^{T} \sum_{\ell=2}^{L} \left( \hat Y_{t\ell}(\textbf{s}_{i}) - Y_{t\ell}(\textbf{s}_{i}) \right)^2}, \\
\text{MAE}_i &= \frac{1}{T (L-1)}	\sum_{t=1}^{T} \sum_{\ell=2}^{L} \left| \hat Y_{t\ell}(\textbf{s}_{i}) - Y_{t\ell}(\textbf{s}_{i}) \right|, \\
\text{CRPS}_i &= \frac{1}{T (L-1)} \sum_{t=1}^{T} \sum_{\ell=2}^{L} \left( \frac{1}{B}\sum_{b=1}^{B} \left|Y_{t\ell}^{(b)}(\textbf{s}_{i}) - Y_{t\ell}(\textbf{s}_{i})\right| - \frac{1}{2B^2} \sum_{b_1=1}^{B}\sum_{b_2=1}^{B} \left|Y_{t\ell}^{(b_1)}(\textbf{s}_{i}) - Y_{t\ell}^{(b_2)}(\textbf{s}_{i})\right| \right), \\
\text{CVG}_i &= \frac{1}{T (L-1)}	\sum_{t=1}^{T} \sum_{\ell=2}^{L} I(L_{t\ell}(\textbf{s}_{i}) \leq Y_{t\ell}(\textbf{s}_{i}) \leq U_{t\ell}(\textbf{s}_{i})),
\end{align*}
where $\hat{Y}_{t\ell}(\textbf{s}_{i}) = \sum_{b=1}^{B} Y_{t\ell}^{(b)}(\textbf{s}_{i})/B$ with $Y_{t\ell}^{(b)}(\textbf{s}_{i})$ the $b$th posterior predictive replicate of $Y_{t\ell}(\textbf{s}_{i})$, from the left-out location $\bs_i$. Also, $(L_{t\ell}(\textbf{s}_{i}), U_{t\ell}(\textbf{s}_{i}))$ is the $90\%$ predictive interval for $Y_{t\ell}(\textbf{s}_{i})$, i.e., the $5$th and $95$th percentiles of the MCMC samples $Y_{t\ell}^{(b)}(\textbf{s}_{i})$ ($b=1,\ldots,B$), and $I(\cdot)$ is the indicator function. The smaller the RMSE, MAE and CRPS values, the better the model performance.  However, the target for CVG is proximity to $0.90$.

\section{Results} \label{sec:analysis}

We summarize, using LOOCV, the comparison of models with differing inclusion of the foregoing spatial GPs.  Each model was fitted to the daily maximum temperature series in months  MJJAS for the 60 years from 1956 to 2015.  Then, we  present  the results for the fitting of the  full model over the study region.

In the MCMC fitting we ran 10 chains, with 200,000 iterations for each chain, to obtain samples from the joint posterior distribution. The first 100,000 samples were discarded as burn-in and the remaining 100,000 samples were thinned to retain 100 samples from each chain for posterior inference.  MCMC diagnostics for the full model are shown in Section~S2.3 of the Supplementary Materials.

\subsection{Validation and model comparison } \label{sec:metrics_results}

The full model considered includes four spatial GPs.  To compare models and assess the importance of the proposed GPs,  simpler models incorporating 0, 1, 2, or 3 GPs are fitted.  $M_p$ with $p = 0,1,\ldots,4$  denotes a model including $p$ spatial processes that are specified  in parentheses.  For example,  $M_1(\beta_0(\bs))$ is the model with a single spatial process for the intercept; for simplicity the full model is denoted $M_4$.

Using the criteria in Section~\ref{sec:metrics} with LOOCV for each of the 18  available locations, Table~\ref{tab:model_comparison} summarizes  the  averages across sites for the four metrics. The strongest improvement in predictive performance  is obtained by adding a spatially varying intercept process, i.e., $M_{1}(\beta_{0}(\bs))$.  The inclusion of  the other GPs does not yield a clear improvement in  performance.  This is not surprising, since the GP for intercepts explicitly rewards predicting the mean and random realizations well in order to agree with the held-out values.
However,  the usefulness of the other GPs with regard to effectively capturing autocorrelations and variances at the observed sites will be seen in Section~\ref{sec:full_model}.

\begin{table}[t]
\caption{Mean value across the 18 sites of the performance metrics for models with different  spatial GPs.}
\label{tab:model_comparison}
\begin{tabular}{lcccc} 
\hline\noalign{\smallskip}
&RMSE&MAE&CRPS&CVG\\
\noalign{\smallskip}\hline\noalign{\smallskip}
$M_0$&4.49&3.64&2.57&0.894\\
$M_1(\beta_0(\textbf{s}))$&4.36&3.53&2.49&0.901\\
$M_1(\alpha(\textbf{s}))$&4.49&3.64&2.57&0.894\\
$M_1(\rho_Y(\textbf{s}))$&4.49&3.64&2.57&0.895\\
$M_1(\sigma_\epsilon(\textbf{s}))$&4.49&3.63&2.56&0.893\\
$M_2(\beta_0(\textbf{s}),\sigma_\epsilon(\textbf{s}))$&4.36&3.53&2.48&0.901\\
$M_3(\beta_0(\textbf{s}),\alpha(\textbf{s}),\sigma_\epsilon(\textbf{s}))$&4.36&3.53&2.49&0.899\\
$M_3(\alpha(\textbf{s}),\rho_Y(\textbf{s}),\sigma_\epsilon(\textbf{s}))$&4.49&3.63&2.56&0.894\\
$M_4$&4.36&3.53&2.48&0.900\\
\noalign{\smallskip}\hline
\end{tabular}
\end{table}

Table~S4 in the Supplementary Materials provides details, by site, for the metrics in Table~\ref{tab:model_comparison}.  The locations with poorest fit for all of the models are Pamplona and Tornos, the only ones with CRPS greater than $3$. They also show large RMSE and MAE as well as poor CVG. For the other locations, the CVG of all the models is closer to the nominal value 0.90.  In particular, $M_4$ not only has the best CVG on average, but the variability of the $\text{CVG}_i$'s with respect to the nominal 0.90 is the lowest of all the models.

\subsection{Results for the full model} \label{sec:full_model}

Here, we show fitted and prediction results for the full model, $M_4$, and demonstrate the  need to include the four GPs. The parameters $\alpha,\beta_1,\beta_2,\beta_3,\alpha(\bs)$, and $\sigma_\alpha$ have been rescaled to interpret them in terms of the original measure of the covariates. Table~\ref{tab:parameters}  summarizes the posterior mean and credible intervals of the model parameters, including standard deviation of random effects.

The harmonic coefficients $\beta_1$ and $\beta_2$  indicate the strong seasonality in the temperature series. The coefficient $\beta_3$ supplies the gradient of temperature corresponding to elevation, approximately $-7^{\circ}$C  per 1,000 m. This value agrees  with the exploratory  analysis in   Section~\ref{sec:data},  and  the  average  environmental lapse rate \citep{navarro2018}. The linear trend coefficient, $\alpha$, indicates that the average increase in temperature is $0.21^{\circ}$C per decade. \cite{pena2021} found a similar trend ($0.27^{\circ}$C per decade) in the summer maximum temperature in Spain (1956--2015).   The posterior  mean of the autoregresive spatial process,  $\rho_Y$,  confirms the strong  serial correlation of  daily temperatures.

The other parameters are standard deviations linked to the spatio-temporal effects of the model.  The posterior mean of  $\sigma_\epsilon$,  the mean of the spatially varying standard deviations of  the pure error process $\epsilon_{t\ell}^{(Y)}(\bs)$, is close to  $3^{\circ}$C. This value doubles  the posterior mean of $\sigma_{\beta_0}$ which represents  the spatial variability of the  mean level $\beta_0(\bs)$, and triples the posterior mean of $\sigma_\lambda$, linked to the  variability of  the yearly random effects $\psi_t$.  The magnitude of the remaining standard deviation parameters is smaller.

\begin{table}[t]
\caption{Posterior mean and $90\%$ credible intervals for the parameters of $M_4$.}
\label{tab:parameters}
\begin{tabular}{ lcc } 
\hline\noalign{\smallskip}
& mean & credible interval \\
\noalign{\smallskip}\hline\noalign{\smallskip}
$\beta_0$ (intercept) & 25.70 & $(24.30, 27.16)$ \\
$\alpha$  (trend)     & $0.0207$ & $(-0.0074,  0.0490)$ \\ 
$\beta_1$ (sine)      & $13.18$ & $(13.00, 13.37)$ \\ 
$\beta_2$ (cosine)    & $0.633$ & $(0.558, 0.709)$ \\ 
$\beta_3$ (elevation) & $-0.0069$ & $(-0.0084, -0.0054)$ \\ 
$\rho_Y$              & $0.691$ & $(0.606, 0.762)$ \\ 
$\sigma_\epsilon$     & $2.963$ & $(2.433, 3.515)$ \\ 
$\sigma_\eta$         & $0.230$ & $(0.201, 0.264)$ \\ 
$\sigma_\lambda$      & $0.936$ & $(0.799, 1.088)$ \\ 
$\sigma_{\beta_0}$    & $1.492$ & $(1.154, 1.939)$ \\ 
$\sigma_{\alpha}$     & $0.0283$ & $(0.0211, 0.0376)$ \\ 
$\sigma_{\rho_Y}$     & $0.339$ & $(0.263, 0.435)$ \\ 
$\sigma_{\sigma_\epsilon^2}$   & $0.404$ & $(0.312, 0.522)$ \\ 
\noalign{\smallskip}\hline
\end{tabular}
\end{table}

\begin{figure}[t]
	\centering
	\includegraphics[width=\textwidth]{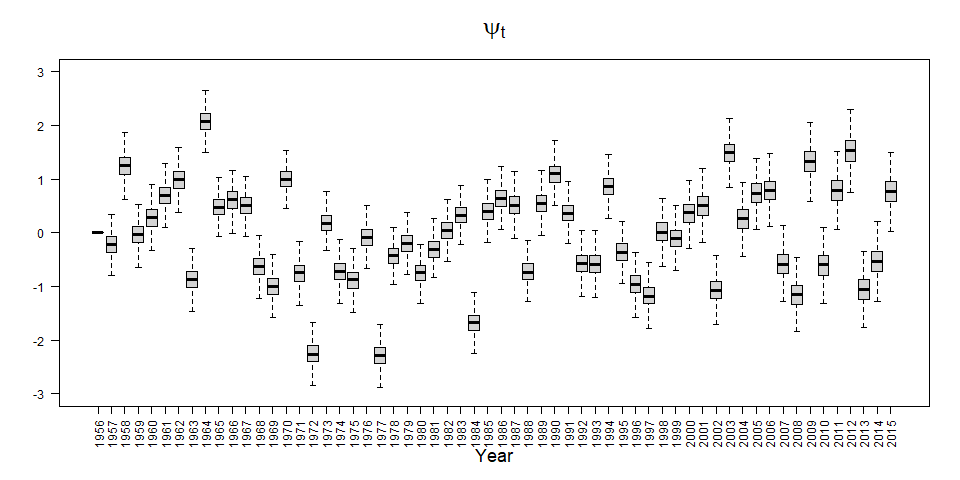}
	\caption{Boxplots of the posterior distributions of the annual random effects $\psi_t$ in $M_4$.}
	\label{fig:psi}
\end{figure}

With $\rho_{\psi}=0$, the yearly random effects, $\psi_t$, are, a priori, distributed as $N(0, \sigma^{2}_{\lambda})$.  The posteriors are summarized using boxplots in Figure~\ref{fig:psi}.   It  is observed   that the effects  may add or subtract in a given year up to roughly $2.5^{\circ}$C, with  a standard deviation close to $1^{\circ}$C. These yearly random effects  are able to capture historical events like the extremely cold summer of 1977 in Spain or the European heat wave in 2003 \citep{pena2021}.

\begin{figure}[p]
	\centering
	\includegraphics[width=.45\textwidth]{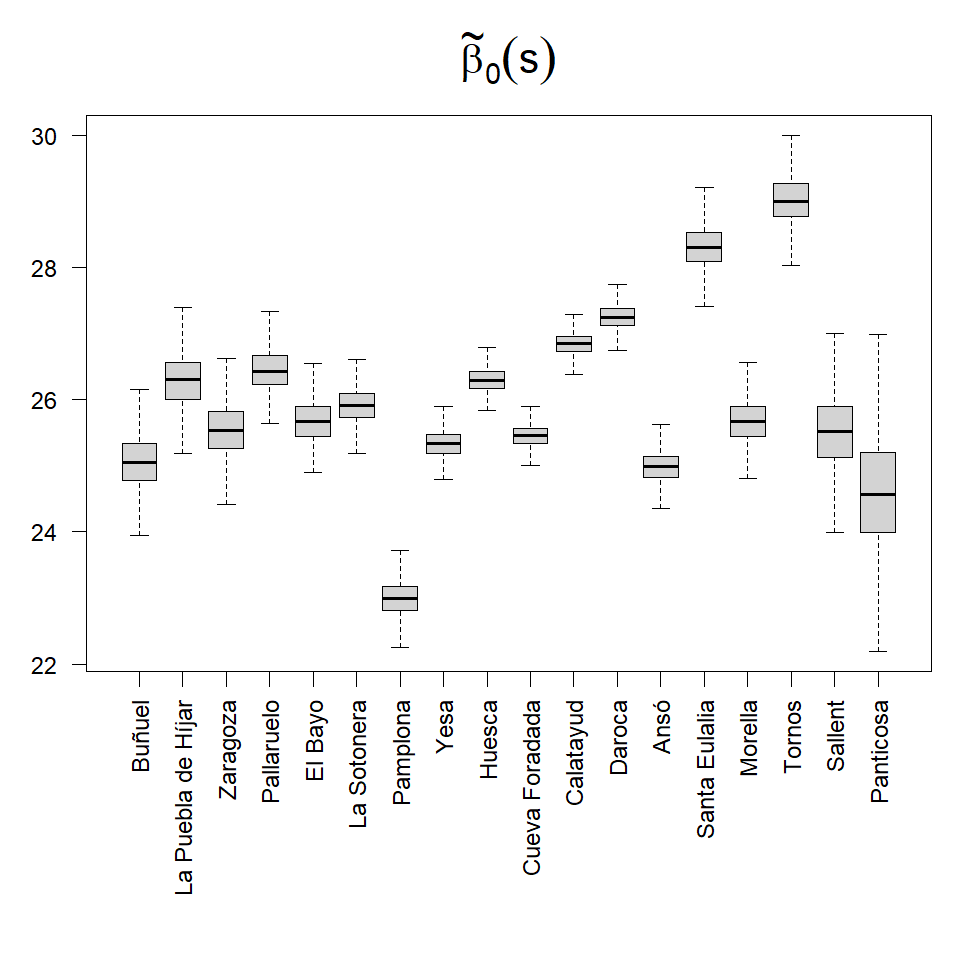}
	\includegraphics[width=.45\textwidth]{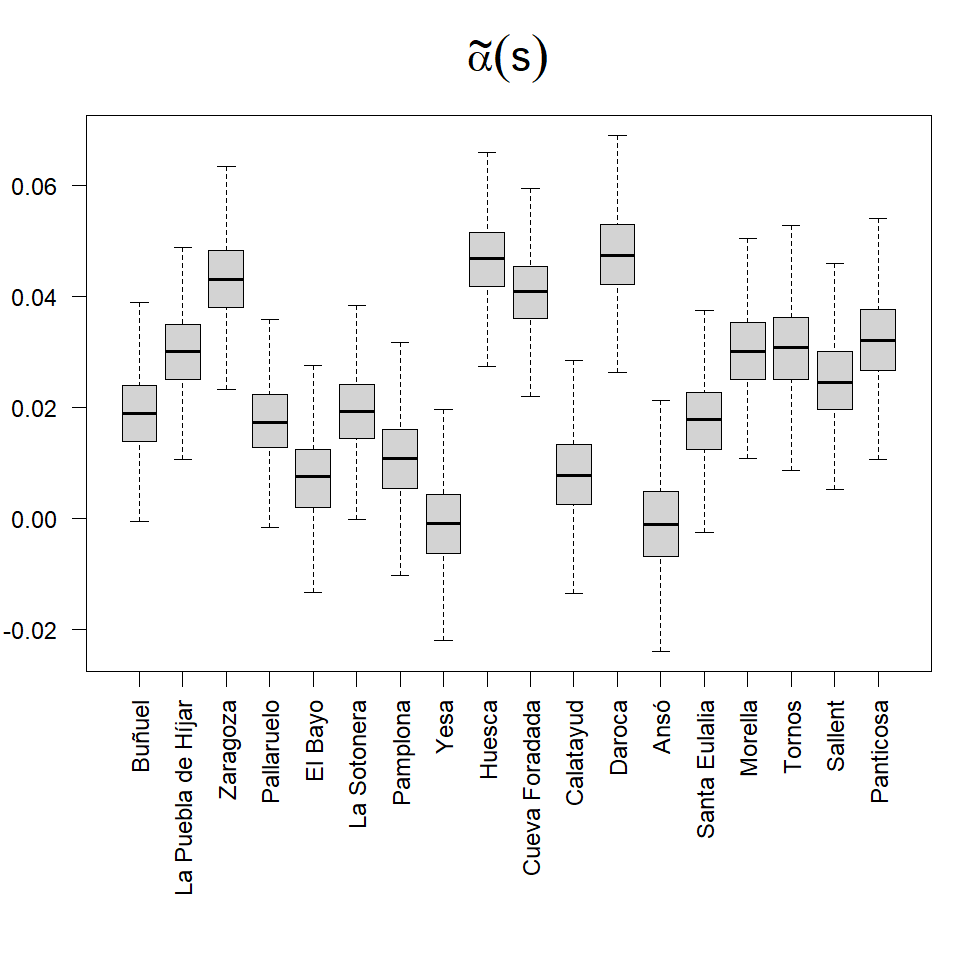}
	\includegraphics[width=.45\textwidth]{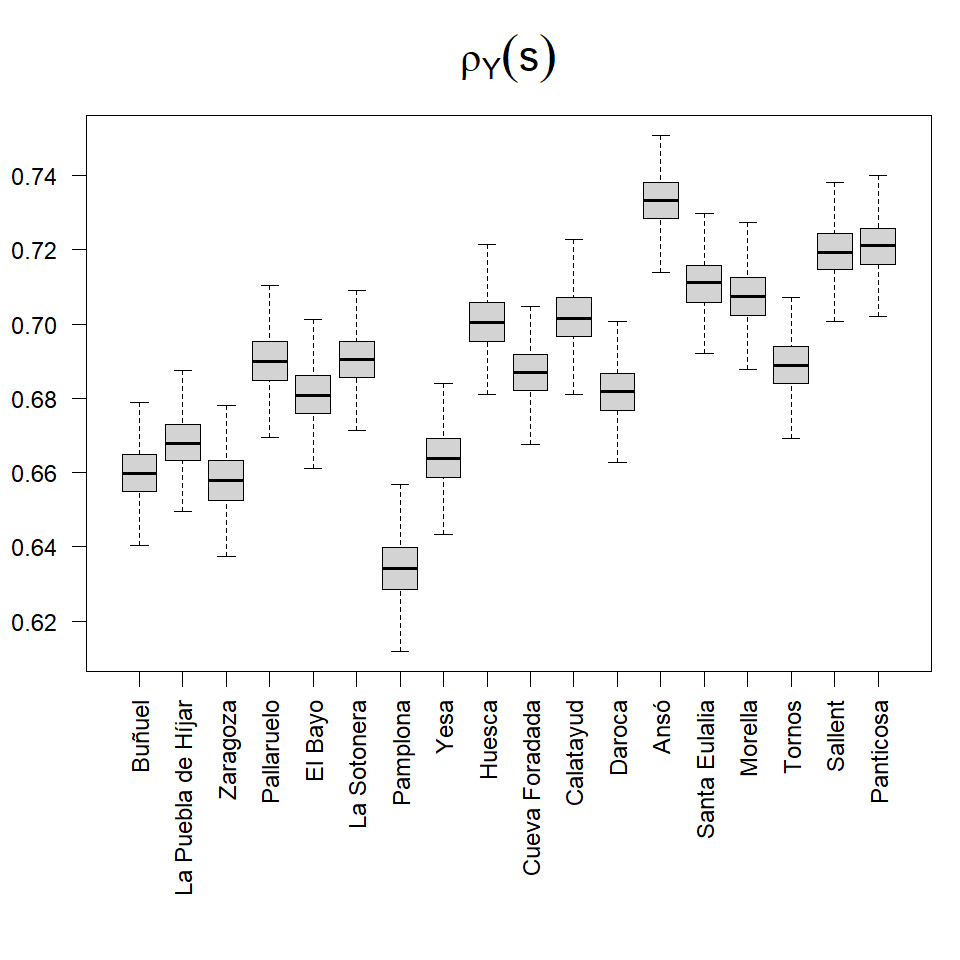}
	\includegraphics[width=.45\textwidth]{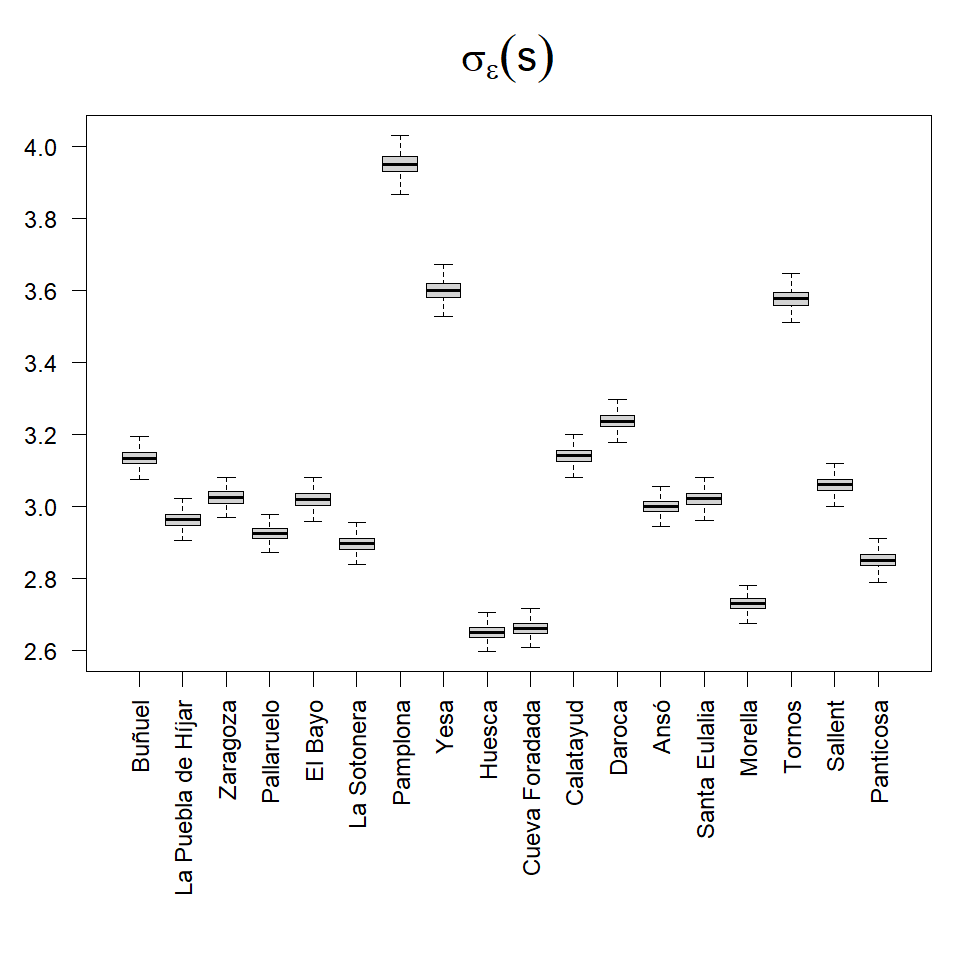}
	\caption{Boxplots of the posterior distributions of the spatial random effects, $\tilde{\beta}_0(\bs),\tilde{\alpha}(\bs),\rho_Y(\bs),\sigma_\epsilon(\bs)$, in $M_4$.  Locations are sorted by elevation, from lowest to highest.}
	\label{fig:GP}
\end{figure}

The posterior distributions at the observed locations of the four spatial processes in $M_4$, $\tilde{\beta}_0(\bs),\tilde{\alpha}(\bs),\rho_Y(\bs)$, and $\sigma_\epsilon(\bs)$,  are summarized in Figure~\ref{fig:GP} using boxplots.  The  boxplots of the locations are sorted from the lowest to the highest elevation in the horizontal axis. They confirm the need to consider the four GPs to represent the great climatic variability of the region under study.  To show the spatial behavior of the spatial processes over the entire region, maps of their posterior means, obtained by a model-based Bayesian kriging, are presented in  Figure~\ref{fig:spatialGP}. In Section~S3.2 of the Supplementary Materials, the parameters of $M_4$ are compared with the parameters of the local models described in Section~S1.1, and both show good agreement.

\begin{figure}[p]
	\centering
	\includegraphics[width=.45\textwidth]{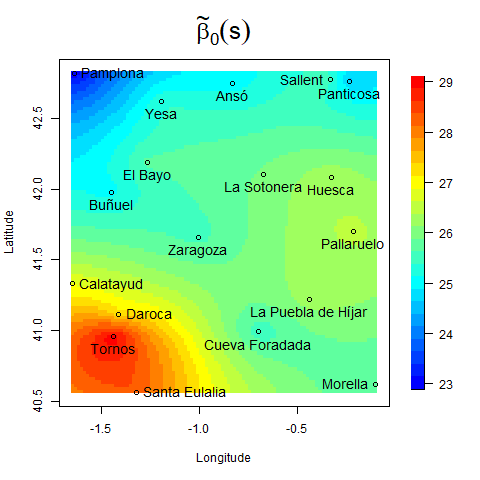}
	\includegraphics[width=.45\textwidth]{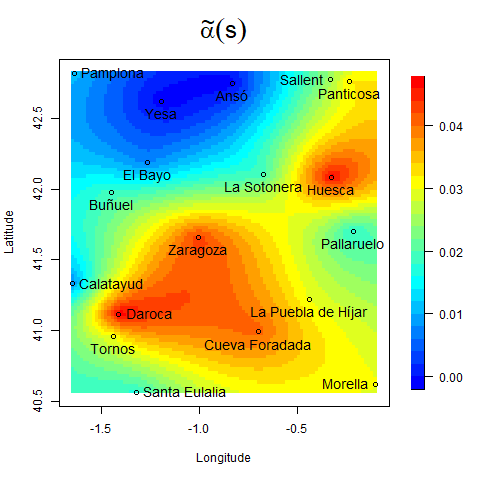}
	\includegraphics[width=.45\textwidth]{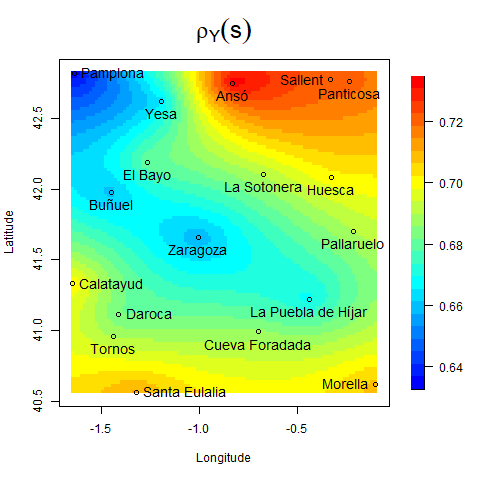}
	\includegraphics[width=.45\textwidth]{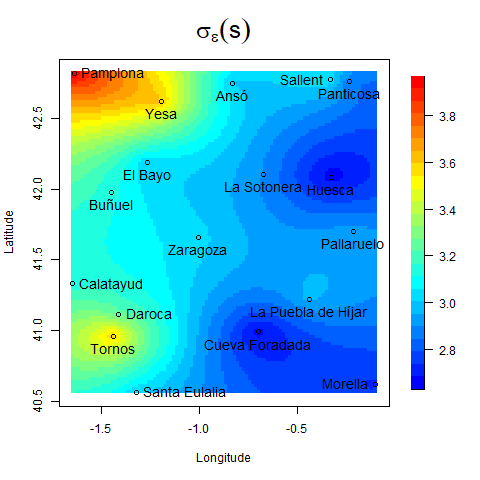}
	\caption{Maps of the  posterior means of the four spatial processes included in $M_4$, obtained by a model-based Bayesian kriging, with resolution $100 \times 100$.}
	\label{fig:spatialGP}
\end{figure}

The top-left plots in Figures~\ref{fig:GP} and \ref{fig:spatialGP} correspond to $\tilde{\beta}_0(\bs)$. The posterior distributions for most of the locations show remarkable differences.  In particular, $\tilde{\beta}_0(\bs)$ has a clear climatic  interpretation. The spatial adjustments provided by this GP help to  improve the fit for the two areas with a similar elevation around 1,000 m but different climates. These areas are the southwest and the north of the region. The former has a warmer climate than the latter,  whose climate is influenced by the proximity of the Atlantic Ocean.

With regard to the spatially varying yearly linear trend, $\tilde{\alpha}(\bs)$,  the top-right plots in Figures~\ref{fig:GP} and  \ref{fig:spatialGP} reveal clear spatial differences in the warming trend. The posterior distributions for higher locations and for the Central Valley are shifted with respect to others. Most of the area shows warming trends, except some  areas in the northwest, e.g., Yesa or Ans{\'o}, whose posterior distributions are centered at zero.

The spatial process for the autoregressive term, $\rho_Y(\bs)$, is clearly necessary in the model. The bottom-left plot in Figure~\ref{fig:GP} shows that the posterior distributions for the 18 locations differ substantially. The posterior means of the $\rho_Y(\bs)$ are positive in all locations and their values seem to have an increasing relation with the elevation.  According to the bottom-left plot in Figure~\ref{fig:spatialGP}, the posterior mean is also related to \emph{cierzo}, a severe northwesterly cold wind that gives rise to a renewal of the atmospheric condition with less warm air masses. This wind reduces the persistence of the temperature and therefore the dependence with respect to the previous day. In the areas affected by \emph{cierzo}, the  mean is around $0.65$, lower than the posterior mean of the mean of $\rho_Y(\bs)$, close to $0.7$.

The need for the $\sigma_\epsilon(\bs)$ process is also clear.  The bottom-right plot in Figure~\ref{fig:GP} reveals strong differences among the posterior distributions of the standard deviations across locations.  The high variability of Pamplona, Yesa and Tornos stands out.  The bottom-right plot in Figure~\ref{fig:spatialGP} confirms the spatial variability of the standard deviation and shows  that higher standard deviations are observed in the  western part of the region.

\subsubsection{Prediction at unobserved locations}
 
Now, we illustrate the use of the full model for prediction at three unobserved sites in the region: Longares (530 m), Olite (390 m), and Guara (800 m). The new sites are marked in red in Figure~\ref{fig:Map} and represent areas with different environmental and climatic characteristics. Longares is located in the  southern half of the region  in a rainfed agricultural area dedicated to the production of wine.  Vines  are  seriously affected by global warming since high temperatures lead to both a decrease in production and a  premature ripening of the grapes. Olite is located in  a rural area in  the northwest where smaller increases in the temperature have been observed;   an incomplete series of observed values is available at this site.  Guara is  an uninhabited area in  the Natural Park  Sierra and Ca{\~n}ones de Guara.  The  prediction of the temperature evolution  in this area is essential to  better understand the changes that have been observed in the ecosystem of the Natural Park.

\begin{figure}[t]
	\centering
\includegraphics[width=0.45\textwidth]{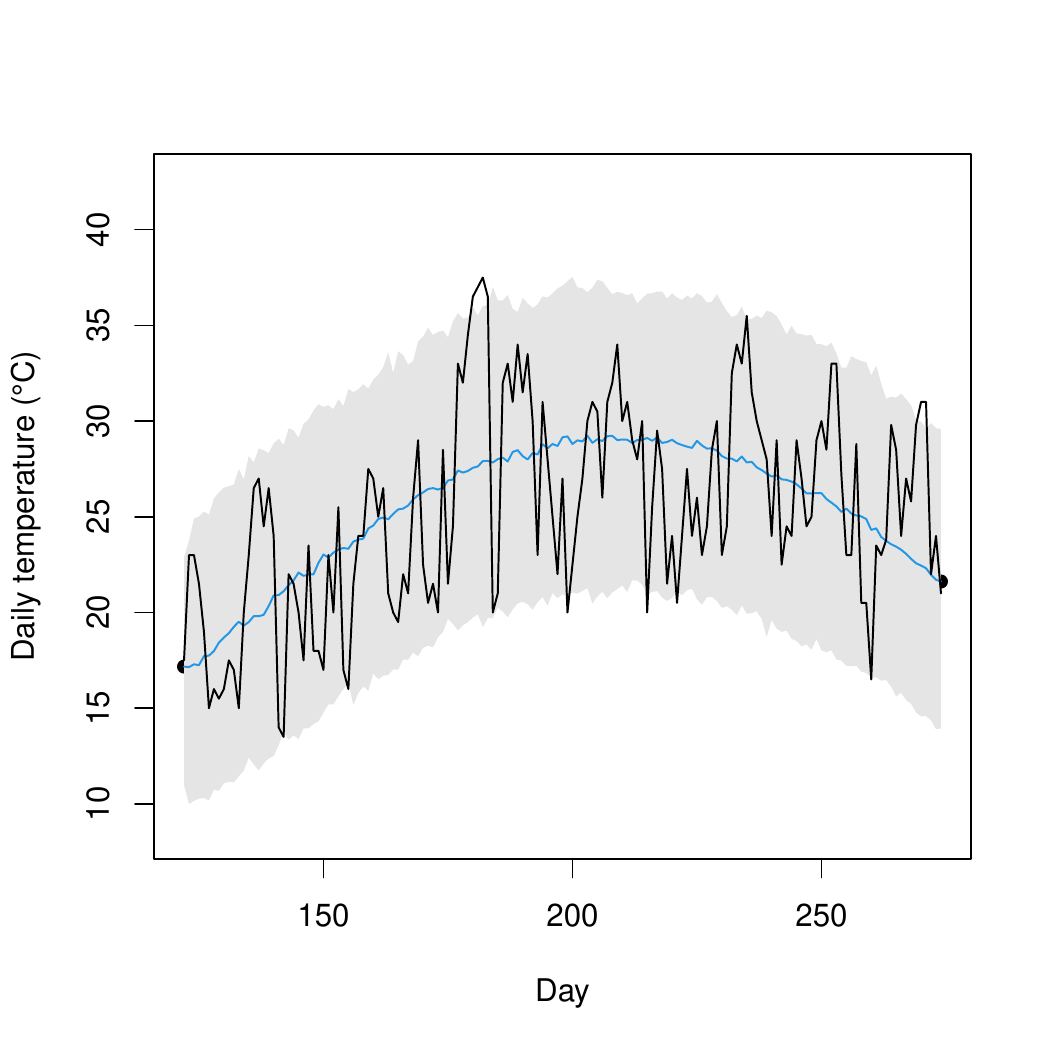}
\includegraphics[width=0.45\textwidth]{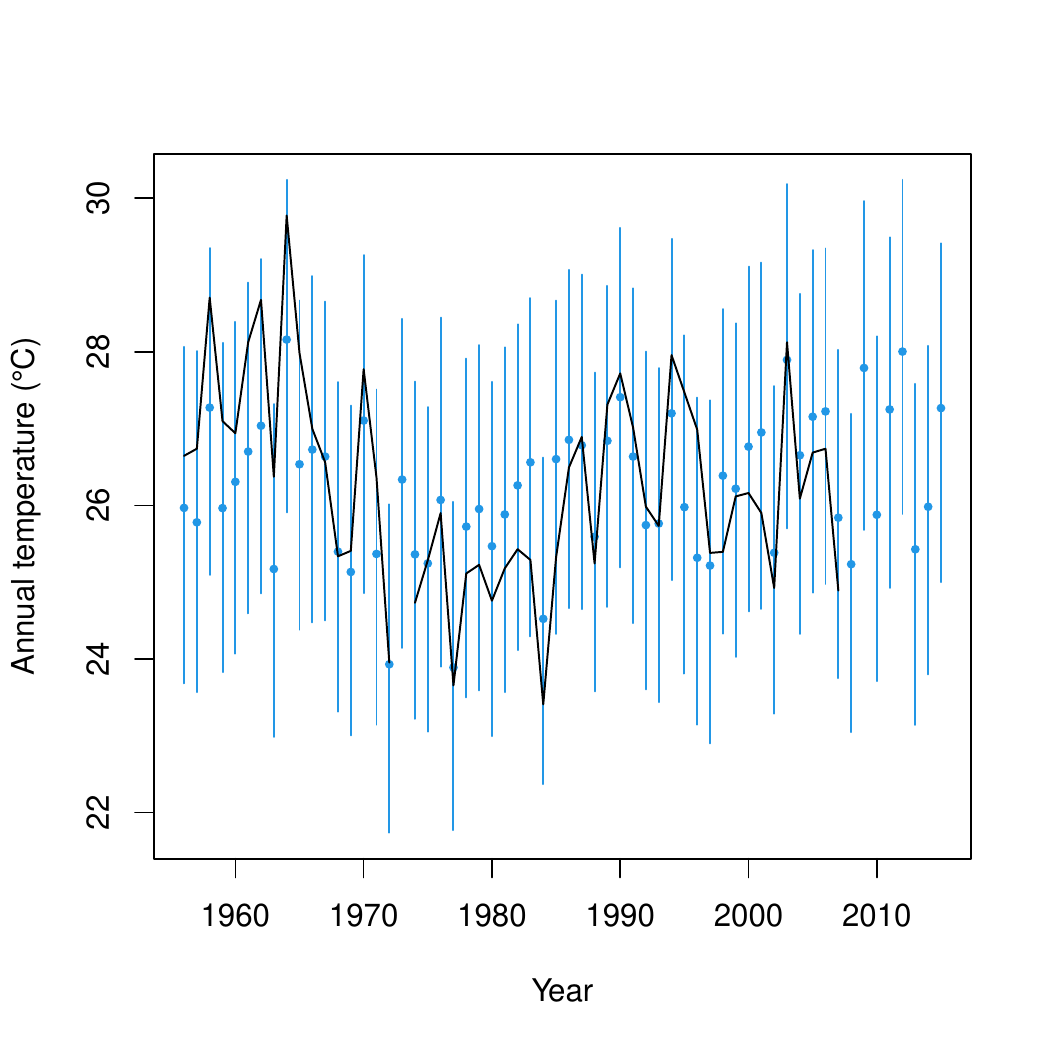}
	\caption{Left: Observed (black rough curve) and posterior predictive means (blue smooth curve) with associated $90\%$ credible intervals of daily maximum temperatures in Olite (1968). Right: Observed yearly averages (black curve) and associated posterior mean and $90\%$ credible intervals.}
	\label{fig:Oli6869}
\end{figure}

We use the model to impute missing values in an observed series using the posterior  predictive  distribution.
Daily temperatures in Olite are available in  the AEMET database from 1968 to 2007, although  with many missing observations.  As an example, Figure~\ref{fig:Oli6869}  shows the plot of the observed series  and the  posterior predictive means with $90\%$ credible intervals for MJJAS days in 1968 and, as a summary, the plot of the observed and the posterior yearly averages with $90\%$ credible intervals. The $90\%$ CVG in the observed data is $92.0\%$.  The agreement between the observed and the predicted data confirms that $M_4$ can be used effectively to impute missing values in Olite.

The posterior distribution of the four spatial processes $\tilde{\beta_0}(\bs)$, $\tilde{\alpha}(\bs)$, $\rho_Y(\bs)$, and $\sigma_{\epsilon}(\bs)$ for the three predicted locations are shown in Figure~S5 of the Supplementary Materials.  The posterior distributions for $\tilde{\beta_0}(\bs)$ in Longares and Guara are similar despite having different elevations. The posterior distributions of $\tilde{\alpha}(\bs)$  in Longares and Guara are very similar while the distribution of Olite is shifted with a posterior mean almost $0.3^{\circ}$C per decade lower. The fitted $\rho_Y(\bs)$'s show the differences in the autocorrelation of temperature in the three locations with posterior means varying from 0.65 to  0.72. The largest differences in the posterior distributions appear in the $\sigma_{\epsilon}(\bs)$.

$M_4$ is also used to evaluate the change over time of the  temperature in the three predicted sites,  using  the posterior predictive distribution of the difference between the  average  in  the 30-year periods 1956--1985 and 1986--2015 (see Figure~S6 in Supplementary Materials). Despite the difference in elevation, the posterior mean of the increment is similar in Longares and Guara, around $1.4^{\circ}$C, while  in Olite it is smaller, $0.5^{\circ}$C and its $90\%$ credible interval $(-0.010, 1.028)$ contains zero. The posterior probability that the  mean in 1986--2015 is higher than in 1956--1985 is 0.94 in Olite and essentially 1 in Longares and Guara.

\section{Summary and future work} \label{sec:summary}

We have proposed a very rich space-time mean model for daily maximum temperatures, fitted over a sixty year period for a region in Spain.  Our specification is continuous in space and autoregressive in time.  In time, autoregression was examined annually and also daily for the summer season within each year.  We find novel spatial structure including spatially varying intercepts and trend coefficients as well as spatially varying autoregression coefficients and variances.

The proposed modeling can be adapted to  other regions, perhaps considering other geographical covariates  such as latitude, longitude or distance to the sea.  Also, the modeling can omit spatial processes that are not necessary, e.g., avoiding $\rho_Y(\bs)$ in a more homogeneous region with a lower variation in elevation. The modeling might also be adapted to other response variables in spatio-temporal problems, such as daily minimum temperature and other environmental variables including daily evapotranspiration or hourly temperature in the sea. The  flexible  autoregression terms can express behavior in  series where serial correlation is an important source of variation.

A limitation of the present analysis is that we have only $18$ monitoring stations so that learning about the spatial surfaces in our modeling is less than we would want. 
Despite this small number of sites, the model has been able to capture the climate variability of the region under study. The spatial random effects  identify areas  with a different mean temperature level, but also areas where the observed warming over time  shows a different trend,  areas where temperature is more persistent (i.e., with a stronger daily serial correlation) or with different variability. The capacity of the fitted model to impute temperature over the entire region allows us to obtain reliable predictions and credible intervals for daily temperature series at unobserved sites. This can be valuable for economical, agricultural or environmental reasons.

Future work will consider different regions providing more available spatial locations $n$. However, the $\mathcal{O}(n^3)$ computational complexity of inverting a $n \times n$ covariance matrix can be prohibitive for implementing the above model for data with large $n$. Reduced rank approximations to GPs may be used to address this computation bottleneck, e.g., Gaussian predictive process \citep{banerjee2008} or nearest-neighbour GP \citep{datta2016}.  As a different challenge, one may wonder whether the low trend values  (blue region) in the top-right plot in Figure~\ref{fig:spatialGP} are actually meaningful. Future work could implement a version of a spatially dependent multiple testing analysis \citep{risser2019} given the posterior draws of $\tilde{\alpha}(\bs)$.  A different future direction will move away from mean modeling to quantile modeling in order to investigate extremes of temperature, both hot and cold.  This will lead to novel development for spatio-temporal quantile regression.

\section*{Acknowledgements}
This work was partially supported by the Ministerio de Ciencia e Innovaci{\'o}n under Grant PID2020-116873GB-I00; Gobierno de Arag{\'o}n under Research Group E46\_20R: Modelos Estoc{\'a}sticos; Jorge Castillo-Mateo was supported by Gobierno de Arag{\'o}n under Doctoral Scholarship ORDEN CUS/581/2020; and Miguel Lafuente was supported by Ministerio de Universidades under Doctoral Scholarship FPU-1505266. The authors thank AEMET for providing the data. The authors are grateful to the Editor, the Associate Editor, and two Referees for their insightful and constructive remarks on an earlier version of the paper.

This version of the article has been accepted for publication at \emph{Journal of Agricultural, Biological and Environmental Statistics}, after peer review but is not the Version of Record and does not reflect post-acceptance improvements, or any
corrections. The Version of Record is available online at: \url{https://doi.org/10.1007/s13253-022-00493-3}.

%
%

\bibliographystyle{spbasic}
\bibliography{ms}

\end{document}


\maketitle

\section{Data and exploratory analysis}

Table~\ref{tab:descrip} shows the elevation in meters for each site. It also summarizes the differences in mean value and standard deviation of the daily maximum temperature in degrees Celsius for each site between both 30-year periods, 1956--1985 and 1986--2015.

\begin{table}[t]
\caption{Elevation, and mean value and standard deviation of the daily maximum temperature for each site in the months MJJAS for the 30-year periods, 1956--1985 and 1986--2015. Difference between means ($\Delta$ mean) and quotient of standard deviations ($Q$ sd) of each period.
\label{tab:descrip} }
\centering
\begin{tabular}{l|ccccc|cc}
\hline\noalign{\smallskip}
\multicolumn{1}{c}{} & \multicolumn{1}{c}{} &\multicolumn{2}{c}{1956--1985} & \multicolumn{2}{c}{1986--2015} &&   \\ 
\multicolumn{1}{c|}{Location} & Elevation & Mean & Sd & Mean & Sd &\multicolumn{1}{c}{$\Delta$ mean}&\multicolumn{1}{c}{$Q$ sd} \\ \noalign{\smallskip}\hline\noalign{\smallskip}
Pamplona               &442&24.2&5.9&25.0&5.9&\phantom{-}0.8&1.0\\ 
Bu{\~n}uel             &242&27.5&5.1&28.5&5.4&\phantom{-}1.0&1.1\\
El Bayo                &360&27.6&5.4&28.0&5.3&\phantom{-}0.4&1.0\\
Morella                &998&22.9&5.2&24.0&5.1&\phantom{-}1.1&1.0\\
Huesca                 &546&26.3&5.1&28.0&5.2&\phantom{-}1.7&1.0\\ 
Tornos                 &1,018&26.1&6.3&27.3&6.3&\phantom{-}1.2&1.0\\ 
Santa Eulalia           &983&25.8&5.7&26.6&5.8&\phantom{-}0.9&1.0\\ 
Calatayud              &600&27.1&5.8&27.6&5.7&\phantom{-}0.5&1.0\\ 
Panticosa              &1,645&17.1&5.4&18.7&5.2&\phantom{-}1.6&1.0\\
La Puebla de H{\'i}jar &245&28.7&5.2&29.8&5.1&\phantom{-}1.1&1.0\\
Ans{\'o}               &860&23.8&5.6&23.6&5.6&-0.2&1.0\\ 
Daroca                 &779&25.6&5.7&27.4&5.7&\phantom{-}1.8&1.0\\ 
Zaragoza               &249&27.6&5.2&29.3&5.2&\phantom{-}1.7&1.0\\
La Sotonera            &413&27.3&5.4&28.1&5.2&\phantom{-}0.8&1.0\\ 
Pallaruelo             &356&28.2&5.6&29.0&5.1&\phantom{-}0.8&0.9\\
Cueva Foradada          &580&25.5&4.7&26.8&5.0&\phantom{-}1.3&1.1\\ 
Sallent &1,285&20.9&5.7&21.8&5.5&\phantom{-}0.9&1.0\\ 
Yesa                   &515&26.3&6.1&26.5&5.7&\phantom{-}0.2&0.9\\ 
\noalign{\smallskip}\hline
\end{tabular}
\end{table}

\subsection{The local model} \label{sec:local}
 
In order to motivate our spatial modeling decisions in Section~3 of the Main Manuscript, in this section we fit independent local models for each location following the steps in that section. However, here we do not center or scale the covariates. The local model for day $\ell$, year $t$ at any location simplifies the full model as
\begin{equation*}
Y_{t\ell} = \mu_{t\ell} + \psi_t + \rho_Y (Y_{t,\ell-1} -  (\mu_{t,\ell-1} + \psi_t)) + \epsilon_{t\ell}^{(Y)},
\end{equation*}
where the fixed effects are
\begin{equation*}
\mu_{t\ell} = \beta_{0} + \alpha t + \beta_{1} \sin(2\pi \ell/365) + \beta_{2} \cos(2 \pi \ell/365).
\end{equation*}
We consider $\psi_{t} \stackrel{iid}{\sim} N(0, \sigma_\lambda^2)$ and $\epsilon_{t\ell}^{(Y)} \stackrel{iid}{\sim} N(0, \sigma_\epsilon^2)$. The interpretation of the model terms is equivalent to that given for the full model in the Main Manuscript, so we do not repeat it here.

\begin{figure}[t]
\begin{center}
\includegraphics[width=\textwidth]{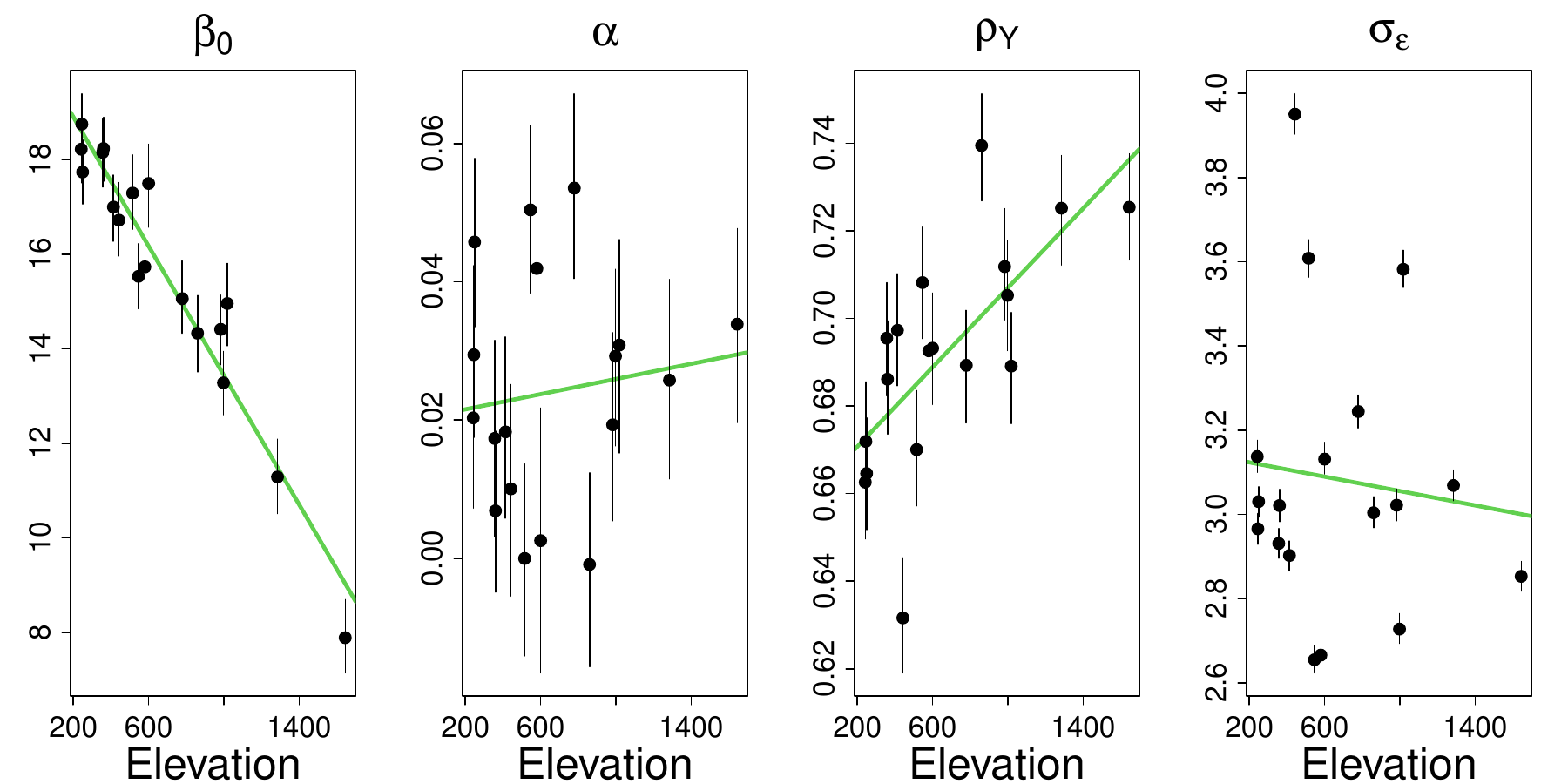}
\caption{Posterior mean and $90\%$ credible interval of the parameters $\beta_0,\alpha,\rho_Y,\sigma_\epsilon$ in the local models. Summaries are sorted by the elevation of the locations. \label{fig:local:model}}
\end{center}
\end{figure}

\begin{figure}[H]
\begin{center}
\includegraphics[width=.45\textwidth]{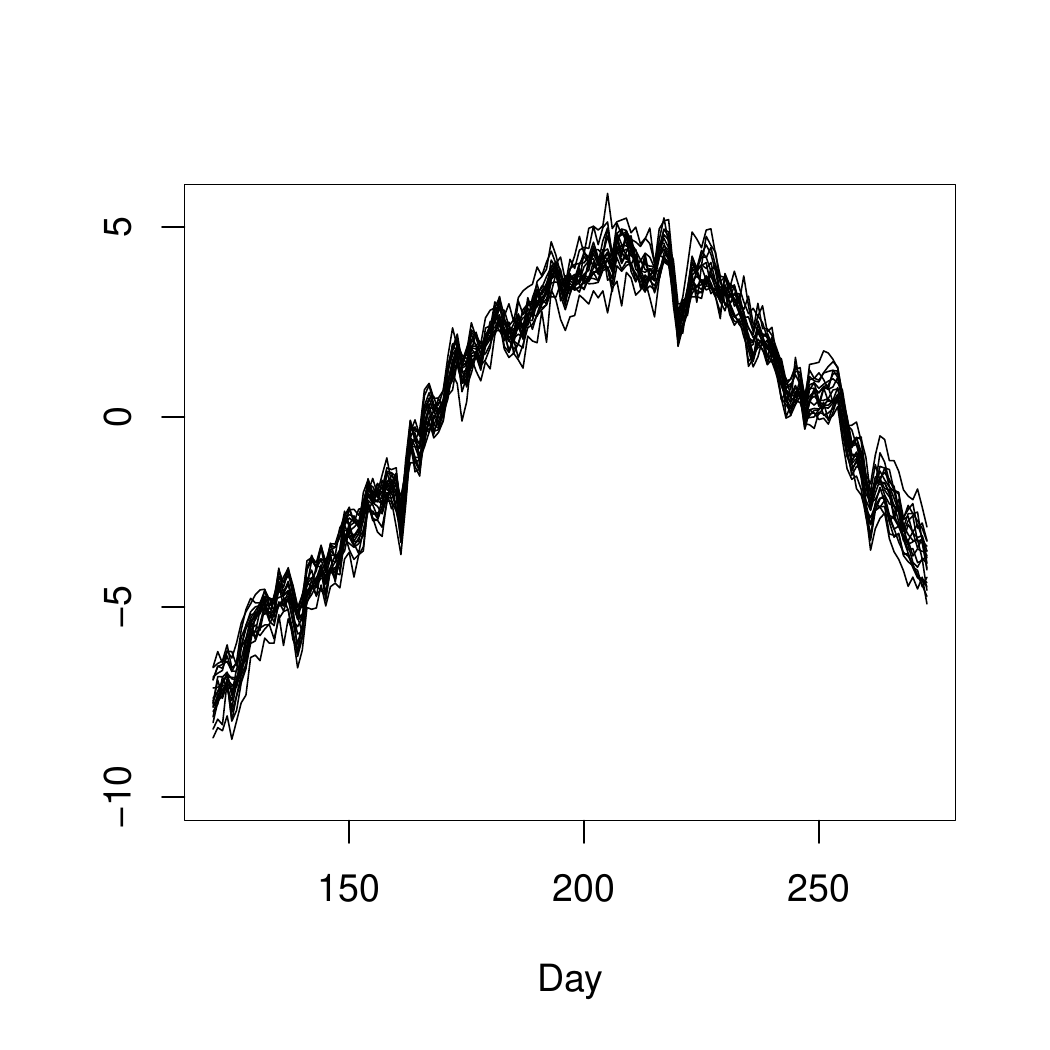}
\includegraphics[width=.45\textwidth]{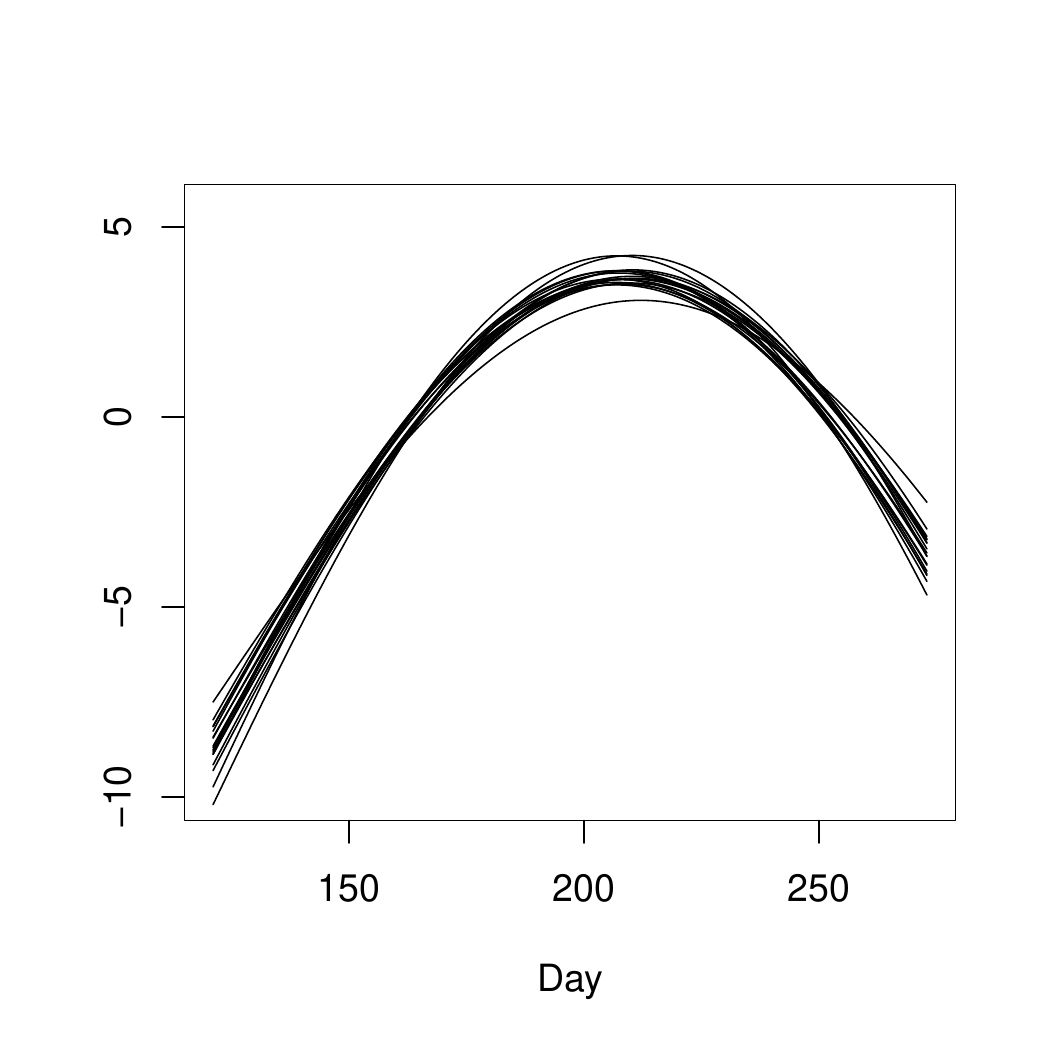}
\caption{Left: Centered empirical mean value of each series across each day of MJJAS. Rigth: Posterior mean of the centered seasonal pattern in the local models.  \label{fig:local:model:seasonal}}
\end{center}
\end{figure}

The parameters are shown in Figure~\ref{fig:local:model} and the seasonal pattern is summarized in \ref{fig:local:model:seasonal}. The first figure shows the posterior mean and $90\%$ credible interval for $\beta_0,\alpha,\rho_Y,\sigma_\epsilon$, for the independently fitted models at each location. Significant differences between locations are observed for the four parameters. The parameter $\sigma_{\lambda}$ did not show a remarkable difference between most locations (not shown).

It is clear that $\beta_0$ is related with elevation, i.e., the temperature is inversely proportional to the elevation of each location. However, this relationship shows considerable noise due to the specific climatic conditions in the southern region, which has stations with an elevation close to 1,000 m with a significantly warmer temperature than stations in the north region with a similar elevation. The spatial variability of $\alpha$ suggests that there is a different warming between locations, that does not seem to be related to elevation. The $\rho_Y$ parameter and the elevation are clearly related, but the variability around the linear relationship cannot be captured by a fixed effect. Finally, the posterior mean of $\sigma_\epsilon$ is quite different between locations with narrow credible intervals, showing a spatial variability of this parameter throughout the region without any relation to elevation.

The left plot in Figure~\ref{fig:local:model:seasonal} shows the empirical seasonal pattern at each location. In particular, the mean value over the years is drawn for each day of MJJAS and each location. The right plot shows the posterior mean of $\beta_{1} \sin(2\pi \ell/365) + \beta_{2} \cos(2 \pi \ell/365)$ for $\ell=1,\ldots,L$. In both plots the patterns have been centered to compare them between locations. In conclusion, the climate of Arag{\'o}n shows a common and unimodal seasonal pattern in MJJAS. In particular, the seasonal component can be characterized by a single harmonic for the entire region.

In summary, this local modeling is useful to find spatial differences and similarities between the full model parameters for different points in space. Furthermore, the inclusion of spatial random effects in the model associated with the intercept, the linear trend, the autoregression coefficient and the variance is justified by these results.

\section{Sampling methods}

\subsection{Gibbs sampling algorithm}
\label{sec:gibbs}

The Bayesian spatio-temporal model can be represented in a hierarchical structure, following \cite{gelfand2012}, we specify distributions for data, process and parameters in three stages,
\begin{align*}
&\text{First stage:} \quad \big[data \mid process,parameters\big]\\
&\text{Second stage:} \quad \big[process \mid parameters\big]\\
&\text{Third stage:} \quad \big[(hyper)parameters\big].
\end{align*}
Although the hierarchical model can be flattened by suitable marginalization, the advantage of the hierarchical structure lies in convenience of specification, ease of interpretation and facilitation of model fitting. In particular, since the model is Gaussian and linear, the Gibbs sampler is expected to be well behaved and convergence to be fairly quick before 50,000 iterations. The hierarchical form leads to the following joint distribution for data, processes and parameters,
\begin{equation}\label{eq:full_model}
\begin{aligned}
&\prod_{i=1}^{n} \prod_{t=1}^{T} \prod_{\ell=2}^{L} \big[Y_{t\ell}(\textbf{s}_i) \mid Y_{t,\ell-1}(\textbf{s}_i), \beta_1,\beta_2,\beta_3,\gamma_t(\bs_i),\rho_Y(\bs_i),\sigma_\epsilon^2(\bs_i)\big] \\
&\prod_{i=1}^{n}\prod_{t=1}^{T}\big[\gamma_t(\textbf{s}_i)  \mid \tilde{\beta}_0(\textbf{s}_i),\tilde{\alpha}(\textbf{s}_i),\psi_t,\sigma_\eta^2\big]
\prod_{t=2}^{T}[\psi_t \mid \psi_{t-1},\rho_\psi,\sigma_\lambda^2]\\
&\big[\{\tilde{\beta}_0(\textbf{s}_i)\} \mid \beta_0,\sigma_{\beta_0}^2,\phi_{\beta_0}\big] 
\big[\{\tilde{\alpha}(\textbf{s}_i)\} \mid \alpha,\sigma_{\alpha}^2,\phi_{\alpha}\big]
\big[\{Z_{\rho_Y}(\textbf{s}_i)\} \mid Z_{\rho_Y},\sigma_{\rho_Y}^2,\phi_{\rho_Y}\big]
\big[\{Z_{\sigma_\epsilon^2}(\textbf{s}_i)\} \mid Z_{\sigma_\epsilon^2},\sigma_{\sigma_\epsilon^2}^2,\phi_{\sigma_\epsilon^2}\big] \\
&\big[\beta_0\big] \big[\alpha\big] \big[\beta_1\big] \big[\beta_2\big] \big[\beta_3\big]
\big[Z_{\rho_Y}\big] \big[Z_{\sigma_\epsilon^2}\big] 
\big[\rho_\psi\big]
\big[\sigma_\lambda^2\big] \big[\sigma_\eta^2\big] \big[\sigma_{\beta_0}^2\big] \big[\sigma_{\alpha}^2\big] \big[\sigma_{\rho_Y}^2\big] \big[\sigma_{\sigma_\epsilon^2}^2\big]\big[\phi_{\beta_0}\big] \big[\phi_{\alpha}\big] \big[\phi_{\rho_Y}\big] \big[\phi_{\sigma_\epsilon^2}\big],
\end{aligned}
\end{equation}
provided one starts any year $t$ with the observed $Y_{t1}(\bs)$.

Defining notation that will be used to shorten the expressions, we denote the elements of the correlation matrices by $\left(r_{jk}^{(\cdot)}\right)^{-1} = R(\phi_\cdot) = \left(\exp\{-\phi_{\cdot} ||\textbf{s}_j - \textbf{s}_k||\}\right)$ where $\cdot$ is any of $\beta_0,\alpha,\rho_Y,\sigma_\epsilon^2$. We denote $X_{t\ell i} = Y_{t\ell}(\textbf{s}_i) - (\mu_{t\ell}(\textbf{s}_i; \btheta_f) + \gamma_t(\bs_i))$ and $X_{t\ell i}^{(-\cdot)} = Y_{t\ell}(\textbf{s}_i) - (\mu_{t\ell}(\textbf{s}_i; \btheta_f) + \gamma_t(\bs_i))^{(-\cdot)}$, where here $\cdot$ is any of $\beta_1,\beta_2,\beta_3,\gamma_t(\bs_i)$ and represents that $\mu_{t\ell}(\textbf{s}_i; \btheta_f) + \gamma_t(\bs_i)$ does not have that component. Finally, we shorten $\sin_\ell = \sin(2\pi\ell/365)$ and $\cos_\ell = \cos(2\pi\ell/365)$. And $a$ and $b$ denote chosen hyperpriors in each case.

The Gibbs sampler algorithm for Equation~\ref{eq:full_model} is initialized giving initial values to all the parameters. Then, updating from iteration $b$ to $b+1$ consists of drawing a sample from the following full conditional distributions:
\begin{itemize}
    \item The full conditional distributions of $\beta_0,\alpha,\beta_1,\beta_2,\beta_3,Z_{\rho_Y},Z_{\sigma_\epsilon^2}$ are all Gaussian, in particular
\begin{align*}
[\beta_0 \mid \cdots] 
&\propto {N}\left( \beta_0 \mid
\frac{\sum_{j,k} r_{jk}^{(\beta_0)} \tilde{\beta}_0(\textbf{s}_k)}{\sum_{j,k} r_{jk}^{(\beta_0)}}, \frac{\sigma_{\beta_0}^2}{\sum_{j,k} r_{jk}^{(\beta_0)}} \right) \times 
{N}(\beta_0 \mid a_{\beta_0}, b_{\beta_0}^2)\\
[\alpha \mid \cdots] 
&\propto {N}\left( \alpha \mid
\frac{\sum_{j,k} r_{jk}^{(\alpha)} \tilde{\alpha}(\textbf{s}_k)}{\sum_{j,k} r_{jk}^{(\alpha)}}, \frac{\sigma_{\alpha}^2}{\sum_{j,k} r_{jk}^{(\alpha)}} \right) \times 
{N}(\alpha \mid a_{\alpha}, b_{\alpha}^2)
\end{align*}
\begin{align*}
[\beta_1 \mid \cdots] 
&\!\begin{multlined}[t][14cm]\propto \prod_{i=1}^n {N}\left(\beta_1 \mid 
\frac{\sum_{t=1}^{T} \sum_{\ell=2}^L (\sin_\ell-\rho_Y(\textbf{s}_i) \sin_{\ell-1}) (X_{t\ell i}^{(-\beta_1)} - \rho_Y(\textbf{s}_i) X_{t,\ell-1,i}^{(-\beta_1)})}{T \sum_{\ell=2}^{L} (\sin_\ell-\rho_Y(\textbf{s}_i) \sin_{\ell-1})^2}, \right.\\
 \left. \frac{\sigma_\epsilon^2(\textbf{s}_i)}{T \sum_{\ell=2}^{L} (\sin_\ell-\rho_Y(\textbf{s}_i) \sin_{\ell-1})^2} \right)
\times {N}(\beta_1 \mid a_{\beta_1}, b_{\beta_1}^2)\end{multlined}\\
[\beta_2 \mid \cdots] 
&\!\begin{multlined}[t][14cm]\propto \prod_{i=1}^n {N}\left(\beta_2 \mid 
\frac{\sum_{t=1}^{T} \sum_{\ell=2}^L (\cos_\ell-\rho_Y(\textbf{s}_i) \cos_{\ell-1}) (X_{t\ell i}^{(-\beta_2)} - \rho_Y(\textbf{s}_i) X_{t,\ell-1,i}^{(-\beta_2)})}{T \sum_{\ell=2}^{L} (\cos_\ell-\rho_Y(\textbf{s}_i) \cos_{\ell-1})^2}, \right.\\
 \left. \frac{\sigma_\epsilon^2(\textbf{s}_i)}{T \sum_{\ell=2}^{L} (\cos_\ell-\rho_Y(\textbf{s}_i) \cos_{\ell-1})^2} \right)
\times {N}(\beta_2 \mid a_{\beta_2}, b_{\beta_2}^2)\end{multlined}\\
[\beta_3 \mid \cdots] 
&\!\begin{multlined}[t][14cm]\propto \prod_{i=1}^n {N}\left( \beta_3 \mid 
\frac{\sum_{t=1}^{T} \sum_{\ell=2}^L (X_{t\ell i}^{(-\beta_3)} - \rho_Y(\textbf{s}_i) X_{t,\ell-1,i}^{(-\beta_3)})}{T (L-1) (1-\rho_Y(\textbf{s}_i)) \text{elev}(\textbf{s}_i)}, \right.\\
 \left. \frac{\sigma_\epsilon^2(\textbf{s}_i)}{T (L-1) (1-\rho_Y(\textbf{s}_i))^2 \text{elev}(\textbf{s}_i)^2} \right)
\times {N}(\beta_3 \mid a_{\beta_3}, b_{\beta_3}^2)\end{multlined}
\end{align*}
\begin{align*}
[Z_{\rho_Y} \mid \cdots] 
&\propto {N}\left( Z_{\rho_Y} \mid 
\frac{\sum_{j,k} r_{jk}^{(\rho_Y)} Z_{\rho_Y}(\textbf{s}_k)}{\sum_{j,k} r_{jk}^{(\rho_Y)}}, \frac{\sigma_{\rho_Y}^2}{\sum_{j,k} r_{jk}^{(\rho_Y)}} \right) \times 
{N}(Z_{\rho_Y} \mid a_{\rho_Y}, b_{\rho_Y}^2)\\
[Z_{\sigma_\epsilon^2} \mid \cdots] 
&\propto {N}\left( Z_{\sigma_\epsilon^2} \mid
\frac{\sum_{j,k} r_{jk}^{(\sigma_\epsilon^2)} Z_{\sigma_\epsilon^2}(\textbf{s}_k)}{\sum_{j,k} r_{jk}^{(\sigma_\epsilon^2)}}, \frac{\sigma_{\sigma_\epsilon^2}^2}{\sum_{j,k} r_{jk}^{(\sigma_\epsilon^2)}} \right) \times 
{N}(Z_{\sigma_\epsilon^2} \mid a_{\sigma_\epsilon^2}, b_{\sigma_\epsilon^2}^2)
\end{align*}
    \item The full conditional distribution of $\rho_\psi$ is a truncated Gaussian distribution within the interval $(a,b)$.
\begin{align*}
[\rho_\psi \mid \cdots] \sim {TN}\left( 
\frac{\sum_{t=2}^T \psi_t \psi_{t-1}}{\sum_{t=2}^T \psi_{t-1}^2},
\frac{\sigma_\lambda^2}{\sum_{t=2}^T \psi_{t-1}^2}, (a_{\rho_\psi},b_{\rho_\psi}) \right)
\end{align*}
    \item The full conditional distributions for $\sigma_\lambda^2,\sigma_\eta^2,\sigma_{\beta_0}^2,\sigma_{\alpha}^2,\sigma_{\rho_Y}^2,\sigma_{\sigma_\epsilon^2}^2$ are all inverse gamma as follows,
\begin{align*}
[1/\sigma_\lambda^2 \mid \cdots] 
&\sim {G}\left(  
\frac{T-1}{2} + a_{\sigma_\lambda}, \frac{1}{2} \sum_{t=2}^T (\psi_t - \rho_\psi \psi_{t-1})^2 + b_{\sigma_\lambda}
\right)\\
[1/\sigma_\eta^2 \mid \cdots] 
&\sim {G}\left(  
\frac{n T}{2} + a_{\sigma_\eta}, \frac{1}{2} \sum_{i=1}^n \sum_{t=1}^T \left(\gamma_t(\textbf{s}_i) - (\tilde{\beta}_0(\textbf{s}_i) + \tilde{\alpha}(\textbf{s}_i) t + \psi_t)\right)^2 + b_{\sigma_\eta}
 \right)\\
[1/\sigma_{\beta_0}^2 \mid \cdots] 
&\sim {G}\left(  
\frac{n}{2} + a_{\sigma_{\beta_0}}, \frac{1}{2} (\{\tilde{\beta}_0(\textbf{s}_i)\} - \beta_0 \bone)^{\top} R(\phi_{\beta_0})^{-1} (\{\tilde{\beta}_0(\textbf{s}_i)\} - \beta_0 \bone) + b_{\sigma_{\beta_0}}
\right)\\
[1/\sigma_{\alpha}^2 \mid \cdots] 
&\sim {G}\left(  
\frac{n}{2} + a_{\sigma_{\alpha}}, \frac{1}{2} (\{\tilde{\alpha}(\textbf{s}_i)\} - \alpha \bone)^{\top} R(\phi_{\alpha})^{-1}(\{\tilde{\alpha}(\textbf{s}_i)\} - \alpha \bone) + b_{\sigma_{\alpha}}
\right)\\
[1/\sigma_{\rho_Y}^2 \mid \cdots] 
&\sim {G}\left(  
\frac{n}{2} + a_{\rho_Y}, \frac{1}{2} (\{Z_{\rho_Y}(\textbf{s}_i)\} - Z_{\rho_Y} \bone)^{\top} R(\phi_{\rho_Y})^{-1} (\{Z_{\rho_Y}(\textbf{s}_i)\} - Z_{\rho_Y} \bone) + b_{\rho_Y}
\right)\\
[1/\sigma_{\sigma_\epsilon^2}^2 \mid \cdots] 
&\sim {G}\left(  
\frac{n}{2} + a_{\sigma_\epsilon^2}, \frac{1}{2} (\{Z_{\sigma_\epsilon^2}(\textbf{s}_i)\} - Z_{\sigma_\epsilon^2} \bone)^{\top} R(\phi_{\sigma_\epsilon^2})^{-1} (\{Z_{\sigma_\epsilon^2}(\textbf{s}_i)\} - Z_{\sigma_\epsilon^2}\bone) + b_{\sigma_\epsilon^2}
\right)\\
\end{align*}
    \item For $\phi_{\beta_0},\phi_{\alpha},\phi_{\rho_Y},\phi_{\sigma_\epsilon^2}$ the simplest solution is to fix the parameter at some reasonable value. An alternative is to discretize the support to say $m$ between 10 to 20 values, obtain and store the collection of $n \times n$ matrices, i.e., inverses and determinants, and then make discrete updates from the following full conditionals.
\begin{align*}
[\phi_{\beta_0} \mid \cdots] &\propto 
|R(\phi_{\beta_0})|^{-1/2} 
\exp\left\{\frac{-1}{2\sigma_{\beta_0}^2} (\{\tilde{\beta}_0(\textbf{s}_i)\} - \beta_0 \bone)^{\top} R(\phi_{\beta_0})^{-1} (\{\tilde{\beta}_0(\textbf{s}_i)\} - \beta_0 \bone) \right\} \\ &\qquad \times U\left\{ a_{\phi_{\beta_0}}^{(1)},\ldots,a_{\phi_{\beta_0}}^{(m)} \right\}\\
[\phi_{\alpha} \mid \cdots] &\propto 
|R(\phi_{\alpha})|^{-1/2} 
\exp\left\{\frac{-1}{2\sigma_{\alpha}^2} (\{\tilde{\alpha}(\textbf{s}_i)\} - \alpha \bone)^{\top} R(\phi_\alpha)^{-1} (\{\tilde{\alpha}(\textbf{s}_i)\} - \alpha \bone) \right\} \\ &\qquad \times U\left\{ a_{\phi_{\alpha}}^{(1)},\ldots,a_{\phi_{\alpha}}^{(m)} \right\}\\
[\phi_{{\rho_Y}} \mid \cdots] &\propto 
|R(\phi_{{\rho_Y}})|^{-1/2} 
\exp\left\{\frac{-1}{2\sigma_{{\rho_Y}}^2} (\{Z_{\rho_Y}(\textbf{s}_i)\} - Z_{\rho_Y} \bone)^{\top} R(\phi_{{\rho_Y}})^{-1} (\{Z_{\rho_Y}(\textbf{s}_i)\} - Z_{\rho_Y} \bone) \right\} \\ &\qquad \times U\left\{ a_{\phi_{{\rho_Y}}}^{(1)},\ldots,a_{\phi_{{\rho_Y}}}^{(m)} \right\}\\
[\phi_{{\sigma_\epsilon^2}} \mid \cdots] &\propto 
|R(\phi_{{\sigma_\epsilon^2}})|^{-1/2} 
\exp\left\{\frac{-1}{2\sigma_{{\sigma_\epsilon^2}}^2} (\{Z_{\sigma_\epsilon^2}(\textbf{s}_i)\} - Z_{\sigma_\epsilon^2} \bone)^{\top} R(\phi_{{\sigma_\epsilon^2}})^{-1} (\{Z_{\sigma_\epsilon^2}(\textbf{s}_i)\} - Z_{\sigma_\epsilon^2} \bone) \right\} \\ &\qquad \times U\left\{ a_{\phi_{{\sigma_\epsilon^2}}}^{(1)},\ldots,a_{\phi_{{\sigma_\epsilon^2}}}^{(m)} \right\}
\end{align*}
    \item The full conditionals for the $\tilde{\beta}_0(\bs_i)$'s and $\tilde{\alpha}(\bs_i)$'s are Gaussian, coming from the joint multivariate Gaussian distributions of $\tilde{\beta}_0(\bs)$ and $\tilde{\alpha}(\bs)$ respectively, and the part of the random effects. Note that we consider the hierarchical centering of these random effects to improve convergence behavior. For $i=1,\ldots,n$, the full conditionals are
\begin{align*}
[\tilde{\beta}_0(\textbf{s}_i) \mid \cdots] 
&\propto {N}\left(  \tilde{\beta}_0(\textbf{s}_i) \mid 
\frac{1}{T} \sum_{t=1}^{T} (\gamma_t(\textbf{s}_i) - \tilde{\alpha}(\textbf{s}_i) t - \psi_t), 
\frac{\sigma_\eta^2}{T}
 \right) \\
&\qquad \times {N}\left( \tilde{\beta}_0(\textbf{s}_i) \mid \beta_0 + 
\frac{\sum_{k \neq i} r_{ik}^{(\beta_0)} (\beta_0 - \tilde{\beta}_0(\textbf{s}_k))}{r_{ii}^{(\beta_0)}},
\frac{\sigma_{\beta_0}^2}{r_{ii}^{(\beta_0)}} \right)\\
[\tilde{\alpha}(\textbf{s}_i) \mid \cdots] 
&\propto {N}\left(  \tilde{\alpha}(\textbf{s}_i) \mid 
\frac{\sum_{t=1}^{T} t (\gamma_t(\textbf{s}_i) - \tilde{\beta}_0(\textbf{s}_i) - \psi_t)}{\sum_{t=1}^T t^2} , 
\frac{\sigma_\eta^2}{\sum_{t=1}^T t^2}
\right) \\
&\qquad \times {N}\left( \tilde{\alpha}(\textbf{s}_i) \mid \alpha + 
\frac{\sum_{k \neq i} r_{ik}^{(\alpha)} (\alpha - \tilde{\alpha}(\textbf{s}_k))}{r_{ii}^{(\alpha)}},
\frac{\sigma_{\alpha}^2}{r_{ii}^{(\alpha)}} \right) 
\end{align*}
    \item The full conditional distributions of the $Z_{\rho_Y}(\bs_i)$'s and $Z_{\sigma_\epsilon^2}(\bs_i)$'s are non-standard. To draw samples from them, we suggest a random walk Metropolis-Hastings algorithm with Gaussian distribution proposals with the mean at the current parameter value. According to \cite{gelman1996}, the variance of the proposals should be tuned until the acceptance rate is between $15\%$ and $40\%$. For $i=1,\ldots,n$, the full conditionals are proportional to
\begin{align*}
[Z_{\rho_Y}(\textbf{s}_i) \mid \cdots] 
&\propto
\exp\left\{ \frac{-1}{2 \sigma_{\epsilon}^2(\textbf{s}_i)} \sum_{t=1}^T \sum_{\ell=2}^L \left( X_{t\ell i} - \frac{e^{Z_{\rho_Y}(\textbf{s}_i)} - 1}{e^{Z_{\rho_Y}(\textbf{s}_i)} + 1} X_{t,\ell-1,i} \right)^2 \right\} \\
&\qquad \times {N}\left( Z_{\rho_Y}(\textbf{s}_i) \mid Z_{\rho_Y} + 
\frac{\sum_{k \neq i} r_{ik}^{(\rho_Y)} (Z_{\rho_Y} - Z_{\rho_Y}(\textbf{s}_k))}{r_{ii}^{(\rho_Y)}},
\frac{\sigma_{\rho_Y}^2}{r_{ii}^{(\rho_Y)}} \right)\\
[Z_{\sigma_\epsilon^2}(\textbf{s}_i) \mid \cdots] 
&\propto
\exp\{Z_{\sigma_\epsilon^2}(\textbf{s}_i)\}^{-T(L-1)/2}
\exp\left\{ \frac{-1}{2 \exp\{Z_{\sigma_\epsilon^2}(\textbf{s}_i)\}} \sum_{t=1}^T \sum_{\ell=2}^L (X_{t\ell i} - \rho_Y(\textbf{s}_i) X_{t,\ell-1,i})^2 \right\} \\
&\qquad \times {N}\left( Z_{\sigma_\epsilon^2}(\textbf{s}_i) \mid Z_{\sigma_\epsilon^2} +
\frac{\sum_{k \neq i} r_{ik}^{(\sigma_\epsilon^2)} (Z_{\sigma_\epsilon^2} - Z_{\sigma_\epsilon^2}(\textbf{s}_k))}{r_{ii}^{(\sigma_\epsilon^2)}},
\frac{\sigma_{\sigma_\epsilon^2}^2}{r_{ii}^{(\sigma_\epsilon^2)}} \right) 
\end{align*}
    \item We obtain the Gaussian full conditionals for the $\psi$'s as follows. For identifiability, $\psi_1$ is fixed to zero. Then, two cases are considered: (i) when $t=2,\ldots,T-1$, and (ii) when $t=T$. Then, respectively
\begin{align*}
[\psi_t \mid \psi_{t-1}, \psi_{t+1}, \cdots] &\propto 
{N}\left( \psi_t \mid 
\frac{1}{n} \sum_{i=1}^n (\gamma_t(\textbf{s}_i) - \tilde{\beta_0}(\textbf{s}_i) - \tilde{\alpha}(\textbf{s}_i) t),
\frac{\sigma_\eta^2}{n}\right) \\
&\qquad \times {N}\left( \psi_t \mid \frac{\rho_\psi (\psi_{t-1} + \psi_{t+1})}{1 + \rho_\psi^2},\frac{\sigma_\lambda^2}{1 + \rho_\psi^2}\right)
\end{align*}
\begin{align*}
[\psi_T \mid \psi_{T-1}, \cdots] &\propto 
{N}\left( \psi_T \mid 
\frac{1}{n} \sum_{i=1}^n (\gamma_T(\textbf{s}_i) - \tilde{\beta_0}(\textbf{s}_i) - \tilde{\alpha}(\textbf{s}_i) T),
\frac{\sigma_\eta^2}{n}\right) \times
{N}\left( \psi_T \mid \rho_\psi \psi_{T-1}, \sigma_\lambda^2\right)
\end{align*}
    \item Finally, the full conditionals for the $\gamma_t(\bs_i)$'s are all Gaussian. For $i=1,\ldots,n$ and $t=1,\ldots,T$, 
\begin{align*}
[\gamma_t(\textbf{s}_i) \mid \cdots] &\propto {N}\left( \gamma_t(\textbf{s}_i) \mid
\frac{\sum_{\ell=2}^L(X_{t\ell i}^{(-\gamma_t(\textbf{s}_i))} - \rho_Y(\textbf{s}_i) X_{t,\ell-1,i}^{(-\gamma_t(\textbf{s}_i))})}{(L-1) (1-\rho_Y(\textbf{s}_i))},
\frac{\sigma_\epsilon^2(\textbf{s}_i)}{(L-1) (1-\rho_Y(\textbf{s}_i))^2}\right) \\ &\qquad \times
{N}\left( \gamma_t(\textbf{s}_i) \mid \tilde{\beta}_0(\textbf{s}_i) + \tilde{\alpha}(\textbf{s}_i)t + \psi_t, \sigma_\eta^2 \right)
\end{align*}
\end{itemize}

Note in the expressions above that the product of Gaussian densities is proportional to a Gaussian density with parameters as follows
\begin{equation*}
\prod_{i=1}^n N\left( x \mid \mu_i, \sigma_i^2 \right) \propto 
N\left( x \mid \sum_{i=1}^{n} \frac{\mu_i}{\sigma_i^2} / \sum_{i=1}^{n} \frac{1}{\sigma_i^2}, 1 / \sum_{i=1}^{n} \frac{1}{\sigma_i^2} \right).
\end{equation*}

\subsection{Composition sampling algorithm and Bayesian kriging}
\label{sec:composition}

Following the notation of Section~3.3 of the Main Manuscript. Once samples of the join posterior distribution have been obtained using the Gibbs sampler algorithm in Section~\ref{sec:gibbs}, one may want to make spatial or space-time predictions. Formally, the posterior predictive distribution for $Y_{t'\ell'}(\bs_0)$ is 
\begin{align*}
\big[ &Y_{t'\ell'}(\bs_0) \mid \bY \big]\\
&=
\int \prod_{\ell = 2}^{\ell'} \big[ Y_{t'\ell}(\bs_0) \mid Y_{t',\ell-1}(\bs_0),\btheta_f,\gamma_{t'}(\bs_0),\rho_Y(\bs_0),\sigma_\epsilon^2(\bs_0) \big]\\ &\qquad\times
\big[\gamma_{t'}(\textbf{s}_0) \mid \tilde{\beta}_0(\textbf{s}_0),\tilde{\alpha}(\textbf{s}_0),\psi_{t'},\sigma_\eta^2\big]\\ &\qquad\times 
\big[\tilde{\beta}_0(\textbf{s}_0) \mid \beta_0,\sigma_{\beta_0}^2,\phi_{\beta_0},\{\tilde{\beta}_0(\textbf{s}_i)\}\big]
\times \big[\tilde{\alpha}(\textbf{s}_0) \mid \alpha,\sigma_{\alpha}^2,\phi_{\alpha},\{\tilde{\alpha}(\textbf{s}_i)\}\big]\\ &\qquad\times 
\big[Z_{\rho_Y}(\textbf{s}_0) \mid Z_{\rho_Y},\sigma_{\rho_Y}^2,\phi_{\rho_Y},\{Z_{\rho_Y}(\textbf{s}_i)\}\big] \times \big[Z_{\sigma_\epsilon^2}(\textbf{s}_0) \mid Z_{\sigma_\epsilon^2},\sigma_{\sigma_\epsilon^2}^2,\phi_{\sigma_\epsilon^2},\{Z_{\sigma_\epsilon^2}(\textbf{s}_i)\}\big]\\ &\qquad\times
\big[ \btheta,\{\tilde{\beta}_0(\textbf{s}_i)\},\{\tilde{\alpha}(\textbf{s}_i)\},\{Z_{\rho_Y}(\textbf{s}_i)\},\{Z_{\sigma_\epsilon^2}(\textbf{s}_i)\},\psi_{t'} \mid \bY \big]\\
&\qquad\times \prod_{\ell = 2}^{\ell'-1}dY_{t'\ell}(\bs_0) d\gamma_{t'}(\bs_0) d\tilde{\beta}_0(\textbf{s}_0) d\tilde{\alpha}(\textbf{s}_0) dZ_{\rho_Y}(\textbf{s}_0) dZ_{\sigma_\epsilon^2}(\textbf{s}_0) d\btheta \\ &\qquad \qquad d\{\tilde{\beta}_0(\textbf{s}_i)\} d\{\tilde{\alpha}(\textbf{s}_i)\} d\{Z_{\rho_Y}(\textbf{s}_i)\} d\{Z_{\sigma_\epsilon^2}(\textbf{s}_i)\} d\psi_{t'},
\end{align*}
where $\bY$ denotes the observed data and $\btheta = (\btheta_f, \btheta_r, Z_{\rho_Y}, Z_{\sigma_\epsilon^2}, \sigma_\eta^2, \sigma_{\rho_Y}^2, \sigma_{\sigma_\epsilon^2}^2, \phi_{\rho_Y}, \phi_{\sigma_\epsilon^2})$ all the parameters. Note that prediction at a new location would require additional modeling for $Y_{t'1}(\bs_0)$, although no additional complications arise. The simplest solution is to obtain these values by ordinary kriging.

The samples from the Gibbs sampler are used to obtain samples from the posterior predictive distribution. First, a random sample is drawn from the posterior distribution using the details in the Gibbs sampler algorithm described in Section~\ref{sec:gibbs}. Then, from each GP, say $W(\bs)$, Bayesian kriging is used to draw a sample from the conditional distribution of $W(\bs_0)$ given $\{W(\bs_i)\}$, details are given in the paragraph below. A sample of $\gamma_{t'}(\bs_0)$ is drawn from
\begin{equation*}
\big[\gamma_{t'}(\bs_0) \mid \cdots \big] \sim N\left(\tilde{\beta}_0(\bs_0) + \tilde{\alpha}(\bs_0) t' + \psi_{t'}, \sigma_\eta^2 \right).
\end{equation*}
Finally, a sample $Y_{t'\ell}(\bs_0)$, $\ell=2,\ldots,\ell'$, is drawn sequentially from the top level model 
\begin{equation*}
\big[ Y_{t'\ell}(\bs_0) \mid \cdots \big] \sim N\left(\mu_{t'\ell}(\bs_0; \btheta_f) + \gamma_{t'}(\bs_0) + \rho_{Y}(\bs_0)(Y_{t',\ell-1}(\bs_0) - (\mu_{t',\ell-1}(\bs_0; \btheta_f) + \gamma_{t'}(\bs_0))), \sigma_\epsilon^2(\bs_0) \right).
\end{equation*}

The details of the Bayesian kriging are as follows. In particular, we are interested in predicting the state of a GP, $W(\bs)$, at a new location $\bs_0$. The joint distribution for $\bs \in \{ \bs_0,\bs_1,\ldots,\bs_n \}$ is a multivariate Gaussian distribution arising from the GP for $W(\bs)$, i.e.,
\begin{equation*}
\begin{pmatrix}
W(\bs_0)\\
\{W(\bs_i)\}
\end{pmatrix} \sim N\left(
\begin{pmatrix}
\mu_0\\
\bmu
\end{pmatrix},  
\begin{pmatrix}
\Sigma_{00}  & \{\Sigma_{i0}\}^\top\\
\{\Sigma_{i0}\} & \bSigma
\end{pmatrix}
\right).
\end{equation*}
Therefore, the conditional distribution of the process at $\bs_0$ is
\begin{equation*}
\big[W(\bs_0) \mid \{W(\bs_i)\} = \bw,\cdots \big] \sim N\left(\mu_0 + \{\Sigma_{i0}\}^\top \bSigma^{-1} (\bw - \bmu), \Sigma_{00} - \{\Sigma_{i0}\}^\top \bSigma^{-1} \{\Sigma_{i0}\}\right),
\end{equation*}
from which we would draw a sample. In particular, to obtain a sample for $\rho_Y(\bs_0)$ or $\sigma_\epsilon^2(\bs_0)$, it is enough to apply to the samples of their associated GPs the inverse function of the transformation applied to them.

\subsection{MCMC convergence diagnostics}

In the MCMC fitting we run 10 chains and 200,000 iterations on each chain to obtain samples from the joint posterior distribution. The first 100,000 samples were discarded as burn-in and the remaining 100,000 samples were thinned (i) to retain 1,000 samples from each chain for computing the estimated potential scale reduction factor \citep[$\hat R$;][]{gelman1992}, and (ii) to retain 100 samples from each chain for computing the effective sample size \citep[ESS;][]{gong2016} out of 1,000 samples and showing trace plots. The samples from (ii) were used for posterior inference.

We check the convergence and mixing of the MCMC algorithm for the full model based on diagnostics and trace plots for all the parameters, although we do not show the individual results for $\psi$'s and $\gamma$'s (a total of $T \times (n + 1)$ parameters). Tables \ref{tab:diag1} and \ref{tab:diag2} show the ESS and the $\hat R$ for the main parameters and the GPs at the observed locations, respectively.
The best ESS should be close to the actual sample size of 1,000, although an ESS of around 200 is considered sufficient. On the other hand, if the 10 chains have converged to the target posterior distribution, then $\hat R$ should be close to 1. In particular, if $\hat R < 1.2$ for all model parameters, one can be confident that convergence has been reached. The ESS is around 1,000 in most parameters, but it is particularly small for $\beta_3$ and $\beta_0(\bs_i)$'s due to the high correlation between them, although their ESS is sufficient. The $\hat R$ is below 1.2 for all the parameters in the tables, $\psi$'s and $\gamma$'s (not shown), which suggests the adequate convergence of the chains. Figure~\ref{fig:trace} shows the trace plots of all 1,000 samples of the parameters.

\begin{table}[!ht]
\caption{Convergence diagnostics for the main parameters of the full model.}
\label{tab:diag1}
\centering
\begin{tabular}{ lcc } 
\hline\noalign{\smallskip}
& ESS & $\hat R$ \\
\noalign{\smallskip}\hline\noalign{\smallskip}
$\beta_0$ & 1,000 & 1.01 \\
$\alpha$  & 1,000 & 1.00 \\ 
$\beta_1$ & 1,000 & 1.00 \\ 
$\beta_2$ & 1,000 & 1.00 \\ 
$\beta_3$ &   226 & 1.12 \\ 
$\rho_Y$  & 1,000 & 1.00 \\ 
$\sigma_\epsilon^2$  & 1,000 & 1.00 \\ 
$\sigma_\eta^2$      & 1,000 & 1.00 \\ 
$\sigma_\lambda^2$   & 1,000 & 1.00 \\ 
$\sigma_{\beta_0}^2$ & 1,000 & 1.00 \\ 
$\sigma_{\alpha}^2$  & 1,000 & 1.00 \\ 
$\sigma_{\rho_Y}^2$  & 1,000 & 1.00 \\ 
$\sigma_{\sigma_\epsilon^2}^2$ & 1,049 & 1.00 \\ 
\noalign{\smallskip}\hline
\end{tabular}
\end{table}

\begin{table}[!ht]
\caption{Convergence diagnostics for the GPs at the observed locations of the full model.}
\label{tab:diag2}
\centering
\begin{tabular}{ lcccccccc } 
\hline\noalign{\smallskip}
\multicolumn{1}{c}{} & \multicolumn{2}{c}{$\tilde{\beta}_0(\bs)$} & \multicolumn{2}{c}{$\tilde{\alpha}(\bs)$} & \multicolumn{2}{c}{$\rho_Y(\bs)$} & \multicolumn{2}{c}{$\sigma_Y^2(\bs)$} \\ 
Location               & ESS & $\hat R$ & ESS & $\hat R$ & ESS & $\hat R$ & ESS & $\hat R$\\
\noalign{\smallskip}\hline\noalign{\smallskip}
Pamplona               & 247 & 1.07 & 1,000 & 1.00 & 1,000 & 1.00 & 1,000 & 1.00 \\
Bu{\~n}uel             & 194 & 1.11 & 1,000 & 1.00 & 1,000 & 1.00 & 1,000 & 1.00 \\ 
El Bayo                & 215 & 1.09 & 1,000 & 1.00 & 1,000 & 1.00 & 1,000 & 1.00 \\ 
Morella                & 190 & 1.10 & 1,000 & 1.00 & 1,000 & 1.00 & 1,000 & 1.00 \\ 
Huesca                 & 429 & 1.04 & 1,000 & 1.00 & 1,000 & 1.00 & 1,000 & 1.00 \\ 
Tornos                 & 219 & 1.10 & 1,000 & 1.00 & 1,000 & 1.00 & 1,000 & 1.00 \\ 
Santa Eulalia          & 213 & 1.10 & 1,000 & 1.00 & 1,000 & 1.00 & 1,000 & 1.00 \\ 
Calatayud              & 745 & 1.02 & 1,000 & 1.00 & 1,000 & 1.00 & 1,120 & 1.00 \\ 
Panticosa              & 232 & 1.12 & 1,000 & 1.00 & 1,000 & 1.00 & 1,000 & 1.00 \\ 
La Puebla de H{\'i}jar & 220 & 1.10 & 1,000 & 1.00 & 1,076 & 1.00 & 1,000 & 1.00 \\ 
Ans{\'o}               & 279 & 1.07 & 1,000 & 1.00 & 1,000 & 1.00 & 1,000 & 1.00 \\ 
Daroca                 & 627 & 1.04 &   915 & 1.00 & 1,000 & 1.00 & 1,000 & 1.00 \\ 
Zaragoza               & 217 & 1.10 & 1,009 & 1.00 & 1,000 & 1.00 &   914 & 1.00 \\ 
La Sotonera            & 247 & 1.08 & 1,000 & 1.00 & 1,000 & 1.00 & 1,000 & 1.00 \\ 
Pallaruelo             & 218 & 1.09 & 1,000 & 1.00 & 1,000 & 1.00 & 1,000 & 1.00 \\ 
Cueva Foradada         & 688 & 1.02 & 1,000 & 1.00 & 1,000 & 1.00 & 1,000 & 1.00 \\ 
Sallent & 181 & 1.12 & 1,000 & 1.00 & 1,000 & 1.00 & 1,000 & 1.00 \\ 
Yesa                   & 334 & 1.05 & 1,000 & 1.00 & 1,000 & 1.00 &   981 & 1.00 \\ 
\noalign{\smallskip}\hline
\end{tabular}
\end{table}

\clearpage

\begin{figure}[H]
\centering
\subfloat[Main parameters (not rescaled)]{\includegraphics[width = 14cm]{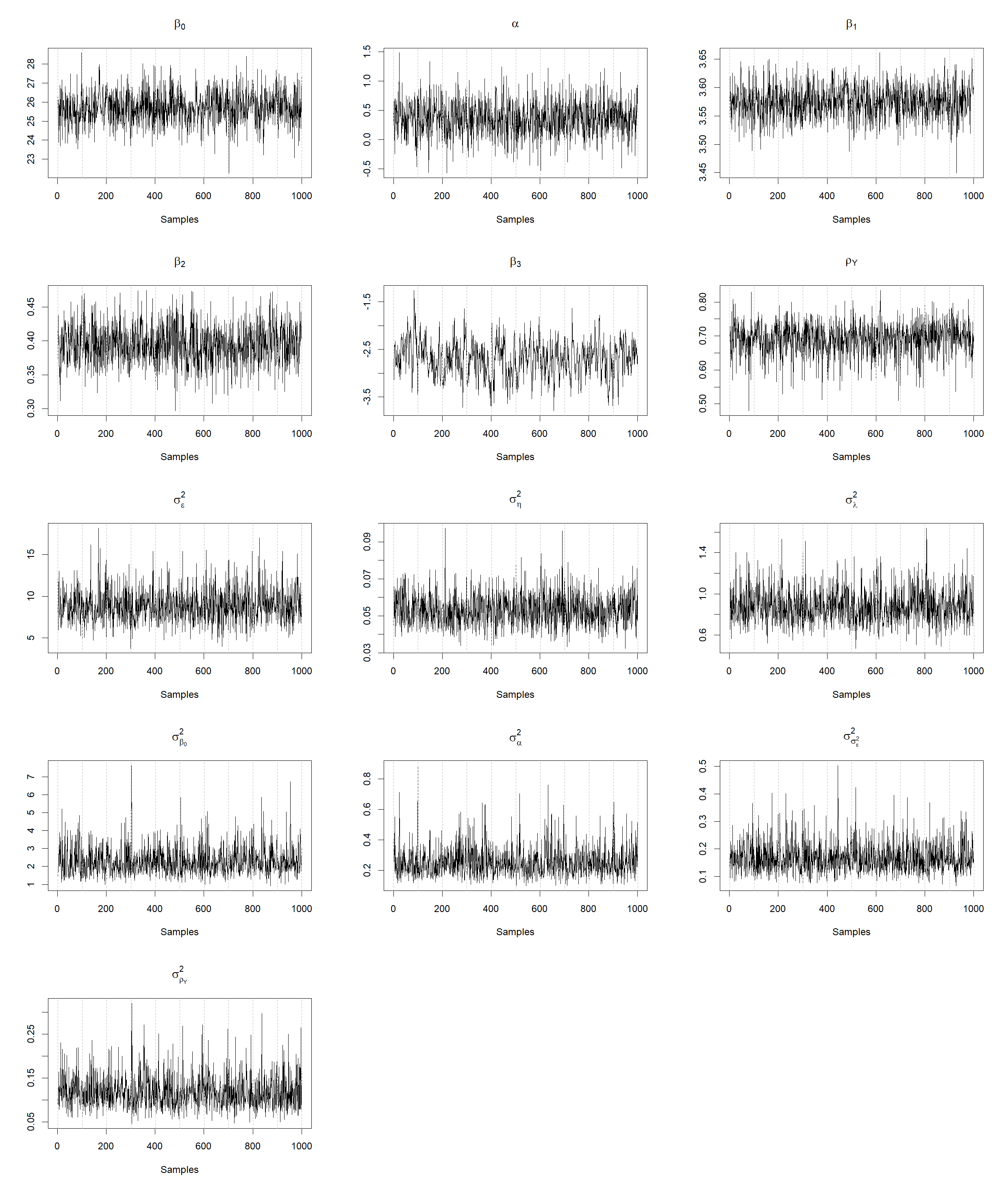}}
\caption[caption]{Trace plots for the parameters (1 of 5). \label{fig:trace}}
\end{figure}

\begin{figure}[H]
\ContinuedFloat
\centering
\subfloat[$\tilde{\beta}_0(\bs)$]{\includegraphics[width = 14cm]{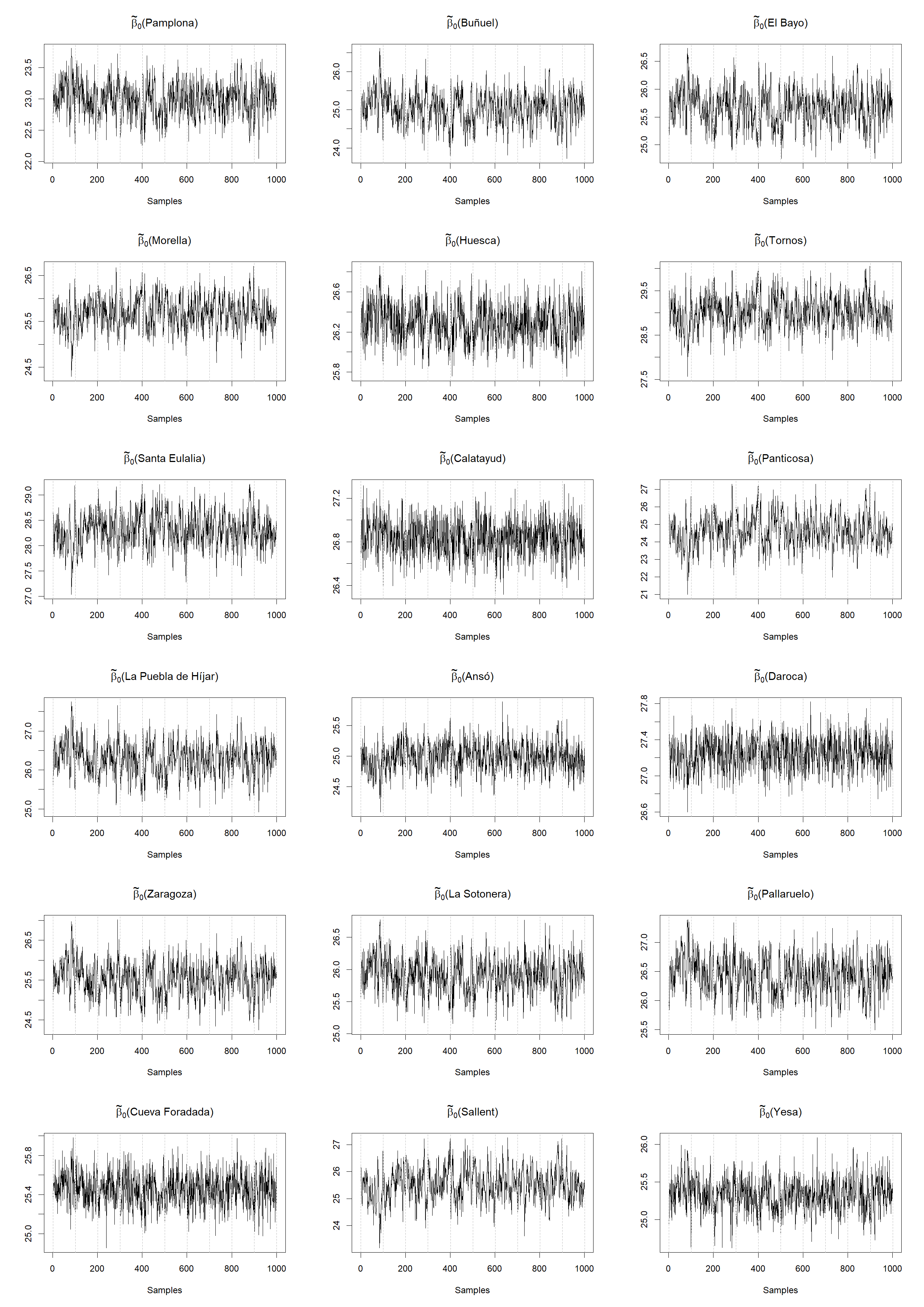}}
\caption[caption]{Trace plots for the parameters (2 of 5).}
\end{figure}

\begin{figure}[H]
\ContinuedFloat
\centering
\subfloat[$\tilde{\alpha}(\bs)$ (not rescaled)]{\includegraphics[width = 14cm]{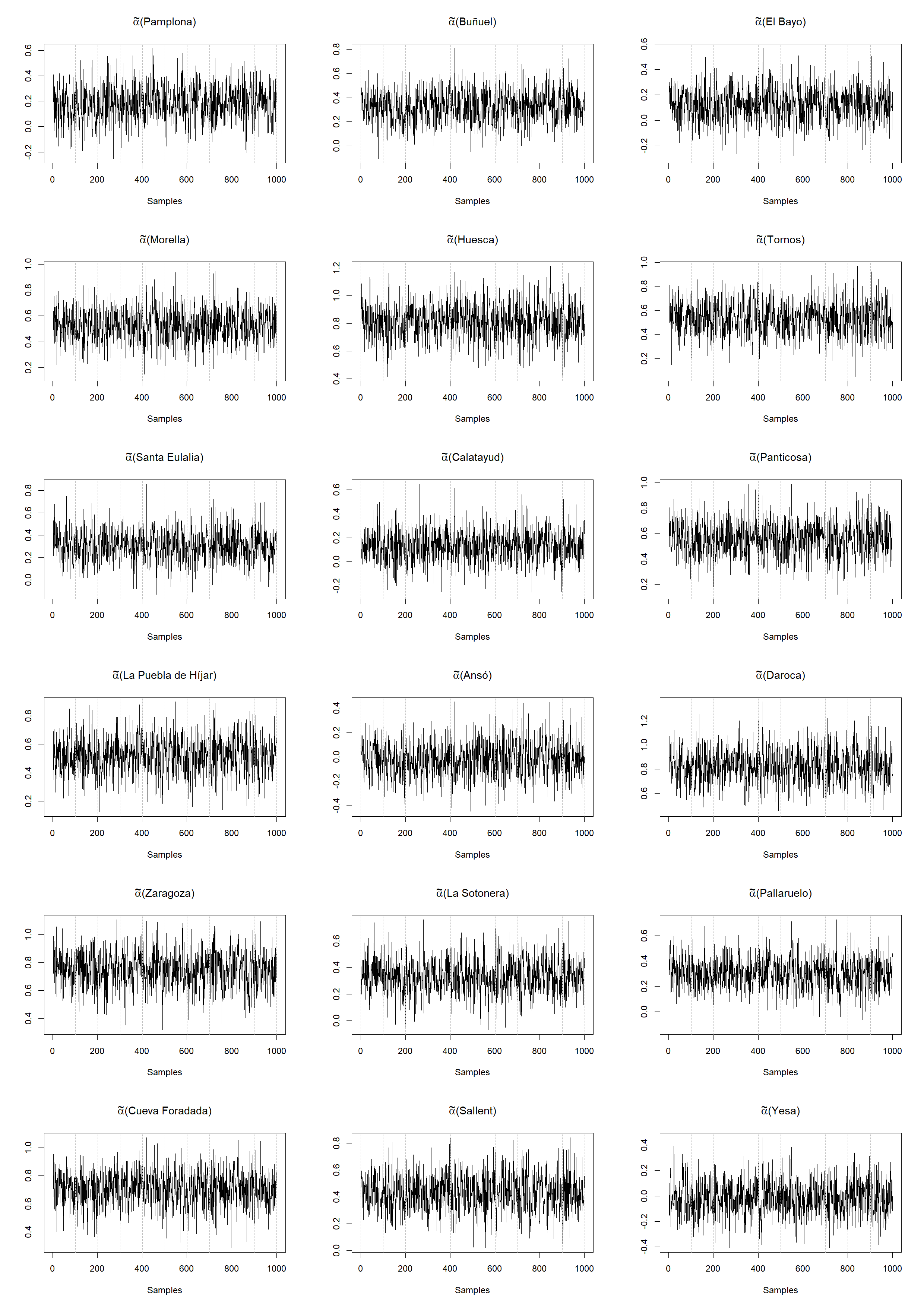}}
\caption[caption]{Trace plots for the parameters (3 of 5).}
\end{figure}

\begin{figure}[H]
\ContinuedFloat
\centering
\subfloat[$\rho_Y(\bs)$]{\includegraphics[width = 14cm]{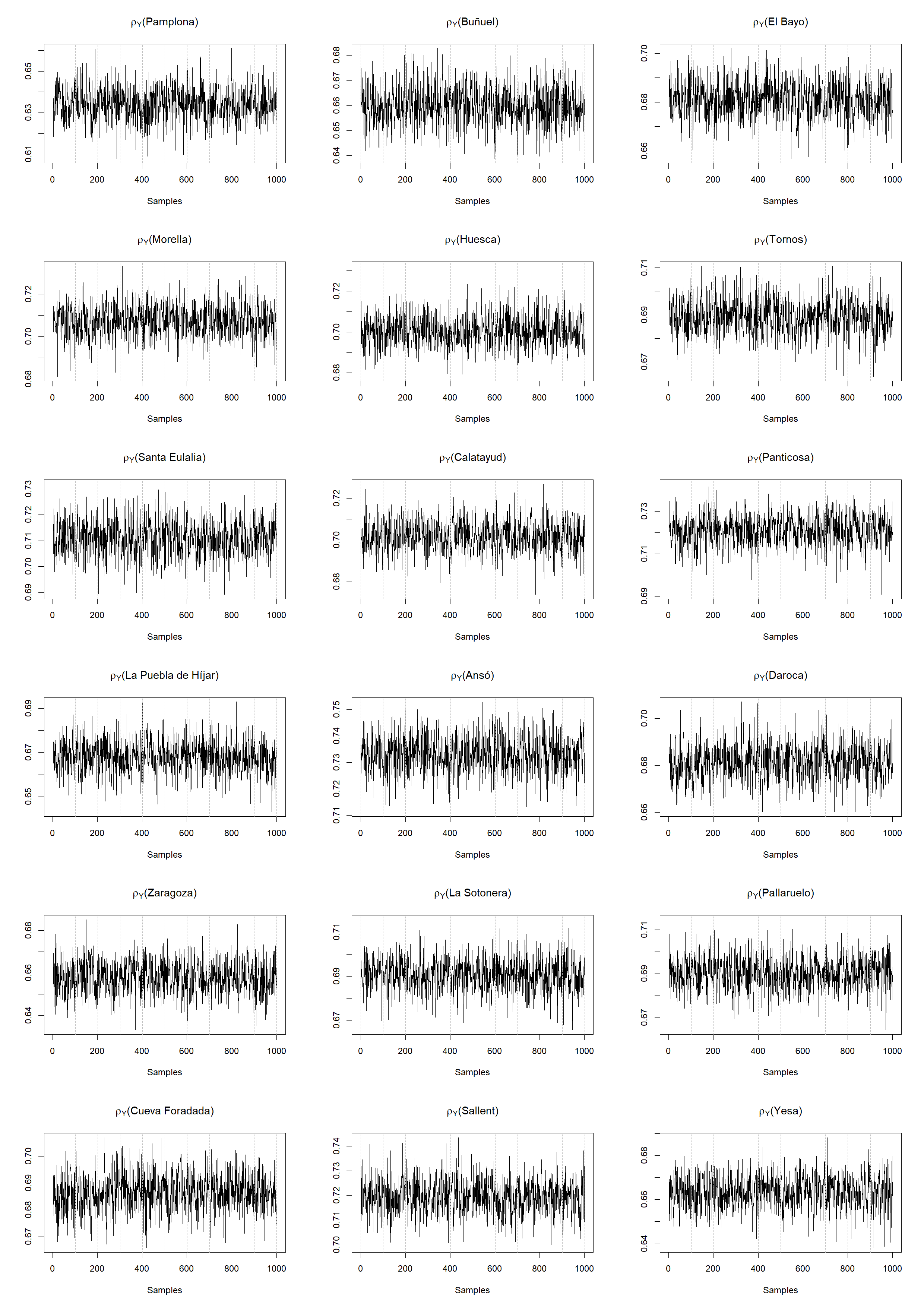}}
\caption[caption]{Trace plots for the parameters (4 of 5).}
\end{figure}

\begin{figure}[H]
\ContinuedFloat
\centering
\subfloat[$\sigma_\epsilon^2(\bs)$]{\includegraphics[width = 14cm]{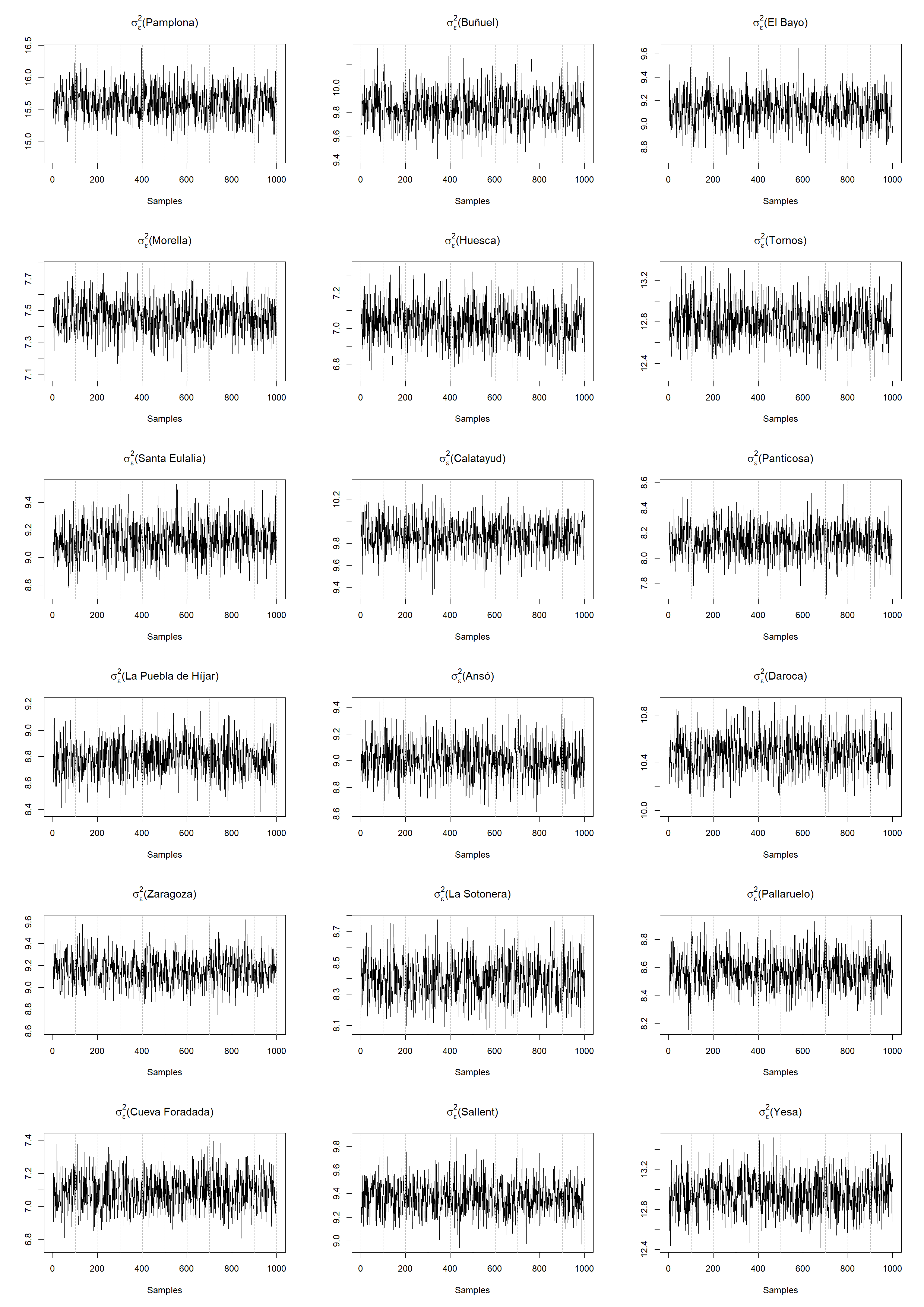}}
\caption[caption]{Trace plots for the parameters (5 of 5).}
\end{figure}

\section{Results}

\subsection{Leave-one-out cross-validation}

The performance of the models is compared using the approach in Section~3.4 of the Main Manuscript based on LOOCV for the 18 locations available. Table~\ref{tab:appendix} summarizes for each site the four considered metrics.

\begin{table}[p]\tiny
	\caption{Value of the performance metrics for models with different spatial GPs for each location. \label{tab:appendix}}
	\begin{center}
		\begin{tabular}{l|cccc|cccc|cccc} 
			Model&\multicolumn{4}{|c|}{Pamplona}&\multicolumn{4}{|c|}{Buñuel}&\multicolumn{4}{|c}{El Bayo}\\
			&RMSE&MAE&CRPS&CVG&RMSE&MAE&CRPS&CVG&RMSE&MAE&CRPS&CVG\\\hline
			$M_0$&5.97&4.89&3.51&0.74&4.25&3.42&2.41&0.92&4.13&3.34&2.35&0.93\\
			$M_1(\beta_0(\textbf{s}))$&5.74&4.72&3.36&0.77&4.21&3.40&2.39&0.92&4.16&3.39&2.37&0.92\\
            $M_1(\alpha(\textbf{s}))$&5.97&4.89&3.51&0.74&4.25&3.43&2.41&0.92&4.12&3.33&2.34&0.93\\
			$M_1(\rho_Y(\textbf{s}))$&5.96&4.88&3.51&0.74&4.24&3.42&2.41&0.92&4.13&3.34&2.34&0.92\\
			$M_1(\sigma_\epsilon(\textbf{s}))$&5.97&4.89&3.48&0.77&4.26&3.43&2.42&0.92&4.13&3.34&2.35&0.94\\
			$M_2(\beta_0(\textbf{s}),\sigma_\epsilon(\textbf{s}))$&5.77&4.74&3.35&0.79&4.21&3.40&2.39&0.92&4.17&3.39&2.37&0.94\\
			$M_3(\beta_0(\textbf{s}),\alpha(\textbf{s}),\sigma_\epsilon(\textbf{s}))$&5.76&4.73&3.35&0.80&4.20&3.40&2.39&0.92&4.15&3.39&2.37&0.94\\
			$M_3(\alpha(\textbf{s}),\rho_Y(\textbf{s}),\sigma_\epsilon(\textbf{s}))$&5.95&4.87&3.48&0.77&4.25&3.43&2.41&0.91&4.12&3.33&2.34&0.93\\
			$M_4$&5.75&4.73&3.36&0.79&4.20&3.40&2.38&0.91&4.15&3.39&2.36&0.93\\
		\end{tabular}
		
		\vspace{2mm}
		
		\begin{tabular}{l|cccc|cccc|cccc} 
			Model&\multicolumn{4}{|c|}{Morella}&\multicolumn{4}{|c|}{Huesca}&\multicolumn{4}{|c}{Tornos}\\
			&RMSE&MAE&CRPS&CVG&RMSE&MAE&CRPS&CVG&RMSE&MAE&CRPS&CVG\\\hline
			$M_0$&3.89&3.05&2.20&0.94&3.75&3.01&2.14&0.95&5.92&4.95&3.50&0.75\\
			$M_1(\beta_0(\textbf{s}))$&3.89&3.06&2.21&0.95&3.75&3.00&2.14&0.95&5.23&4.34&3.04&0.81\\
			$M_1(\alpha(\textbf{s}))$&3.88&3.05&2.20&0.94&3.76&3.02&2.15&0.95&5.92&4.95&3.50&0.75\\
			$M_1(\rho_Y(\textbf{s}))$&3.89&3.05&2.20&0.94&3.76&3.02&2.15&0.96&5.92&4.96&3.50&0.76\\
			$M_1(\sigma_\epsilon(\textbf{s}))$&3.88&3.05&2.19&0.92&3.75&3.01&2.13&0.94&5.92&4.95&3.49&0.77\\
			$M_2(\beta_0(\textbf{s}),\sigma_\epsilon(\textbf{s}))$&3.87&3.05&2.19&0.94&3.75&3.01&2.13&0.93&5.22&4.33&3.03&0.83\\
			$M_3(\beta_0(\textbf{s}),\alpha(\textbf{s}),\sigma_\epsilon(\textbf{s}))$&3.91&3.07&2.21&0.93&3.75&3.01&2.13&0.93&5.23&4.34&3.03&0.83\\
			$M_3(\alpha(\textbf{s}),\rho_Y(\textbf{s}),\sigma_\epsilon(\textbf{s}))$&3.88&3.05&2.19&0.92&3.76&3.02&2.13&0.94&5.92&4.96&3.49&0.78\\
			$M_4$&3.90&3.06&2.20&0.94&3.75&3.01&2.13&0.93&5.23&4.34&3.03&0.83\\
		\end{tabular}
		
		\vspace{2mm}
		\begin{tabular}{l|cccc|cccc|cccc} 
			Model&\multicolumn{4}{|c|}{Santa Eulalia}&\multicolumn{4}{|c|}{Calatayud}&\multicolumn{4}{|c}{Panticosa}\\
			&RMSE&MAE&CRPS&CVG&RMSE&MAE&CRPS&CVG&RMSE&MAE&CRPS&CVG\\\hline
			$M_0$&4.93&4.11&2.84&0.87&4.53&3.71&2.59&0.90&5.15&4.05&2.91&0.83\\
			$M_1(\beta_0(\textbf{s}))$&4.38&3.56&2.49&0.91&4.42&3.60&2.52&0.90&4.44&3.49&2.50&0.88\\			
			$M_1(\alpha(\textbf{s}))$&4.94&4.11&2.85&0.87&4.55&3.72&2.59&0.90&5.15&4.05&2.91&0.83\\
			$M_1(\rho_Y(\textbf{s}))$&4.93&4.10&2.84&0.87&4.54&3.71&2.59&0.89&5.20&4.09&2.94&0.85\\
			$M_1(\sigma_\epsilon(\textbf{s}))$&4.96&4.13&2.85&0.88&4.53&3.71&2.58&0.91&5.06&3.97&2.86&0.83\\
			$M_2(\beta_0(\textbf{s}),\sigma_\epsilon(\textbf{s}))$&4.38&3.56&2.49&0.92&4.43&3.60&2.52&0.92&4.44&3.49&2.50&0.88\\
			$M_3(\beta_0(\textbf{s}),\alpha(\textbf{s}),\sigma_\epsilon(\textbf{s}))$&4.38&3.57&2.49&0.92&4.45&3.61&2.53&0.91&4.48&3.51&2.52&0.87\\
			$M_3(\alpha(\textbf{s}),\rho_Y(\textbf{s}),\sigma_\epsilon(\textbf{s}))$&4.96&4.13&2.85&0.89&4.55&3.72&2.60&0.91&5.09&4.00&2.88&0.84\\
			$M_4$&4.38&3.57&2.49&0.92&4.45&3.62&2.53&0.91&4.46&3.50&2.51&0.89\\
		\end{tabular}
		
		\vspace{2mm}
		\begin{tabular}{l|cccc|cccc|cccc} 
			Model&\multicolumn{4}{|c|}{La Puebla de Híjar}&\multicolumn{4}{|c|}{Ansó}&\multicolumn{4}{|c}{Daroca}\\
			&RMSE&MAE&CRPS&CVG&RMSE&MAE&CRPS&CVG&RMSE&MAE&CRPS&CVG\\\hline
			$M_0$&4.01&3.24&2.28&0.94&4.56&3.63&2.57&0.89&4.61&3.83&2.65&0.90\\
			$M_1(\beta_0(\textbf{s}))$&4.02&3.24&2.28&0.93&4.44&3.58&2.52&0.90&4.50&3.64&2.56&0.88\\
			$M_1(\alpha(\textbf{s}))$&4.01&3.23&2.28&0.94&4.55&3.63&2.57&0.89&4.61&3.82&2.65&0.89\\
			$M_1(\rho_Y(\textbf{s}))$&4.02&3.25&2.28&0.94&4.56&3.63&2.58&0.89&4.61&3.83&2.65&0.90\\
			$M_1(\sigma_\epsilon(\textbf{s}))$&4.00&3.23&2.26&0.91&4.56&3.63&2.58&0.91&4.61&3.83&2.64&0.93\\
			$M_2(\beta_0(\textbf{s}),\sigma_\epsilon(\textbf{s}))$&4.03&3.25&2.28&0.90&4.45&3.58&2.52&0.92&4.49&3.64&2.55&0.92\\
			$M_3(\beta_0(\textbf{s}),\alpha(\textbf{s}),\sigma_\epsilon(\textbf{s}))$&4.01&3.23&2.27&0.90&4.43&3.58&2.51&0.92&4.50&3.64&2.55&0.92\\
			$M_3(\alpha(\textbf{s}),\rho_Y(\textbf{s}),\sigma_\epsilon(\textbf{s}))$&4.01&3.23&2.27&0.90&4.54&3.62&2.57&0.91&4.61&3.83&2.64&0.93\\
			$M_4$&4.03&3.25&2.28&0.90&4.43&3.58&2.51&0.92&4.50&3.64&2.55&0.92\\
		\end{tabular}
		
		\vspace{2mm}
		\begin{tabular}{l|cccc|cccc|cccc} 
			Model&\multicolumn{4}{|c|}{Zaragoza}&\multicolumn{4}{|c|}{La Sotonera}&\multicolumn{4}{|c}{Pallaruelo}\\
			&RMSE&MAE&CRPS&CVG&RMSE&MAE&CRPS&CVG&RMSE&MAE&CRPS&CVG\\\hline
			$M_0$&4.03&3.28&2.29&0.94&4.00&3.21&2.27&0.94&4.09&3.30&2.33&0.93\\
			$M_1(\beta_0(\textbf{s}))$&4.04&3.28&2.30&0.93&4.00&3.20&2.27&0.93&4.04&3.24&2.29&0.93\\
            $M_1(\alpha(\textbf{s}))$&4.02&3.27&2.30&0.94&4.01&3.21&2.28&0.94&4.11&3.30&2.33&0.93\\
			$M_1(\rho_Y(\textbf{s}))$&4.02&3.28&2.29&0.94&4.00&3.21&2.28&0.94&4.10&3.31&2.33&0.93\\
			$M_1(\sigma_\epsilon(\textbf{s}))$&4.03&3.28&2.29&0.93&4.00&3.21&2.27&0.91&4.08&3.29&2.31&0.90\\
			$M_2(\beta_0(\textbf{s}),\sigma_\epsilon(\textbf{s}))$&4.04&3.28&2.30&0.93&4.00&3.20&2.26&0.90&4.04&3.24&2.29&0.91\\
			$M_3(\beta_0(\textbf{s}),\alpha(\textbf{s}),\sigma_\epsilon(\textbf{s}))$&4.04&3.28&2.30&0.92&4.00&3.21&2.26&0.90&4.06&3.25&2.29&0.90\\
			$M_3(\alpha(\textbf{s}),\rho_Y(\textbf{s}),\sigma_\epsilon(\textbf{s}))$&4.02&3.27&2.29&0.92&4.00&3.21&2.27&0.91&4.10&3.30&2.32&0.90\\
			$M_4$&4.04&3.28&2.30&0.92&4.00&3.21&2.27&0.91&4.06&3.25&2.30&0.90\\
		\end{tabular}
		
		\vspace{2mm}
		\begin{tabular}{l|cccc|cccc|cccc} 
			Model&\multicolumn{4}{|c|}{Cueva Foradada}&\multicolumn{4}{|c|}{Sallent}&\multicolumn{4}{|c}{Yesa}\\
			&RMSE&MAE&CRPS&CVG&RMSE&MAE&CRPS&CVG&RMSE&MAE&CRPS&CVG\\\hline
			${M}_0$&3.69&2.94&2.11&0.95&4.47&3.56&2.52&0.90&4.85&3.96&2.78&0.86\\
			${M}_1(\beta_0(\textbf{s}))$&3.85&3.05&2.18&0.94&4.48&3.67&2.56&0.89&4.86&4.03&2.80&0.86\\
			${M}_1(\alpha(\textbf{s}))$&3.68&2.93&2.10&0.96&4.47&3.56&2.52&0.90&4.83&3.95&2.76&0.87\\
			${M}_1(\rho_Y(\textbf{s}))$&3.69&2.94&2.11&0.95&4.49&3.57&2.53&0.92&4.85&3.96&2.78&0.86\\
			${M}_1(\sigma_\epsilon(\textbf{s}))$&3.69&2.94&2.10&0.95&4.46&3.56&2.52&0.87&4.86&3.96&2.77&0.89\\
			${M}_2(\beta_0(\textbf{s}),\sigma_\epsilon(\textbf{s}))$&3.85&3.05&2.18&0.93&4.48&3.67&2.57&0.86&4.87&4.03&2.79&0.88\\
			${M}_3(\beta_0(\textbf{s}),\alpha(\textbf{s}),\sigma_\epsilon(\textbf{s}))$&3.84&3.04&2.17&0.93&4.48&3.67&2.56&0.86&4.84&4.02&2.78&0.89\\
			${M}_3(\alpha(\textbf{s}),\rho_Y(\textbf{s}),\sigma_\epsilon(\textbf{s}))$&3.68&2.93&2.09&0.95&4.46&3.56&2.52&0.89&4.83&3.95&2.76&0.89\\
			${M}_4$&3.84&3.03&2.17&0.93&4.48&3.66&2.55&0.88&4.85&4.03&2.79&0.88\\
		\end{tabular}
	\end{center}
\end{table}

\subsection{Comparison between local models and the full model}

\begin{figure}[t]
	\centering
	\includegraphics[width=.5\textwidth]{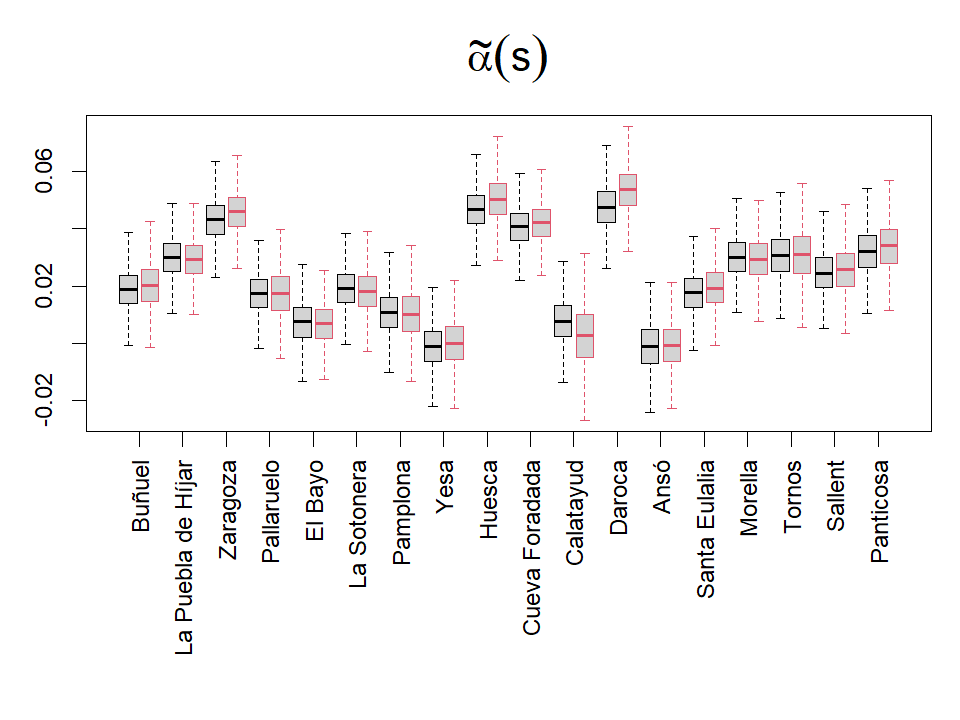}
	\includegraphics[width=.5\textwidth]{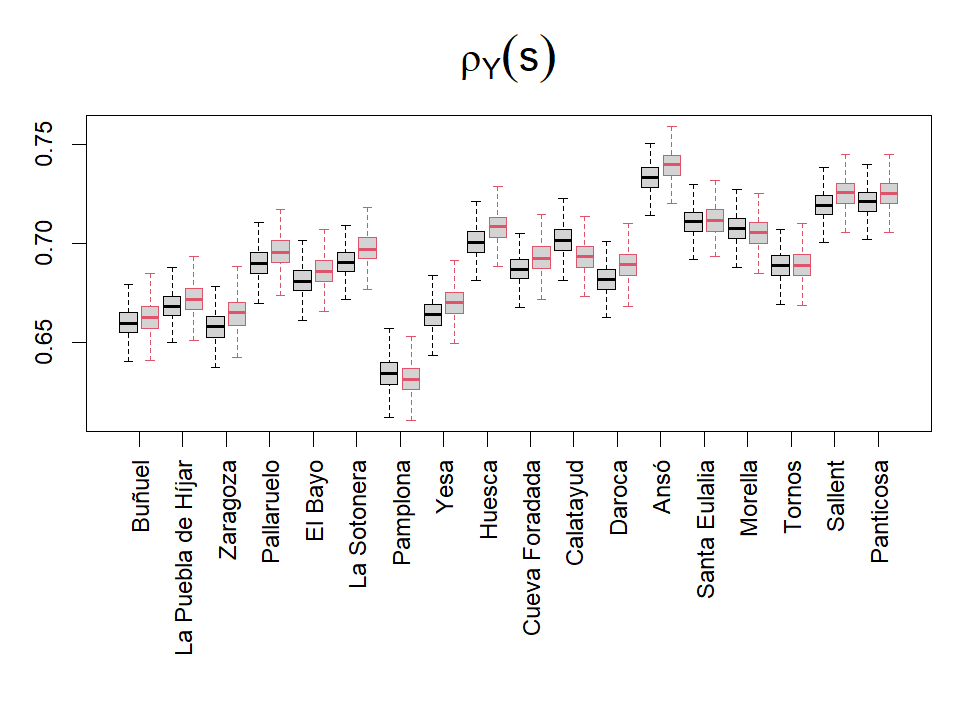}
	\includegraphics[width=.5\textwidth]{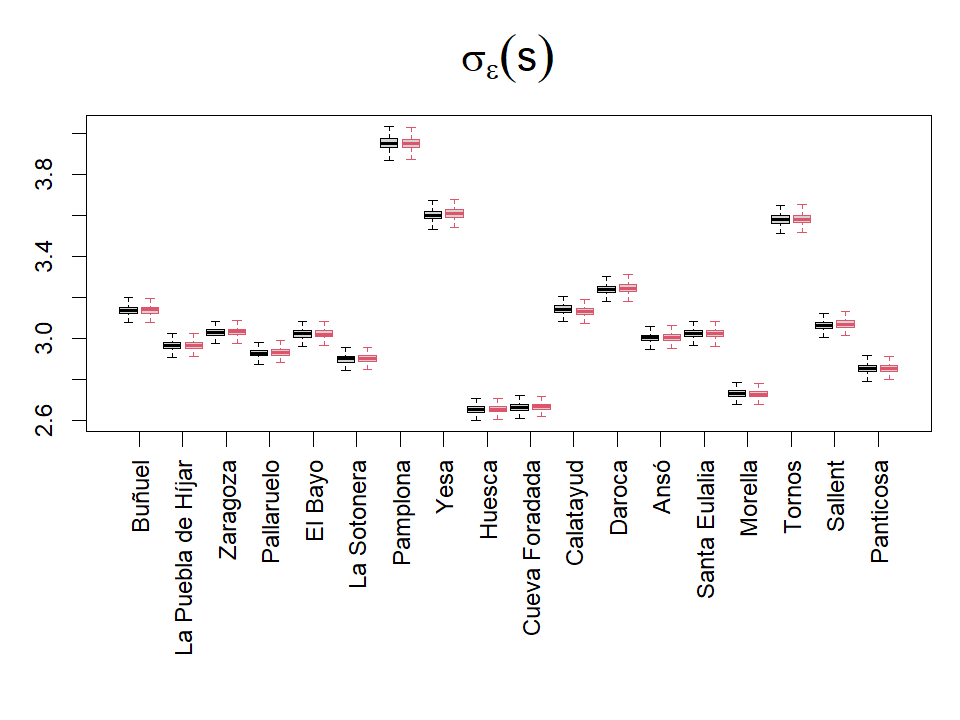}
	\caption{Boxplots of the posterior distributions of the spatial random effects in $M_4$ (black) and of the corresponding coefficients in the local models (red).  Locations are sorted by elevation, from lowest to highest.}
	\label{fig:M4loc}
\end{figure}

Figure~\ref{fig:M4loc} shows a comparison between the posterior distributions at the observed locations of the spatial processes $\tilde{\alpha}(\bs),\rho_Y(\bs)$, and $\sigma_\epsilon(\bs)$, in $M_4$ (black), and the posterior distribution of the corresponding parameters in the local models shown in Section~\ref{sec:local} (red). The results for $M_4$ show a good agreement with results of the local models. This agreement shows that $M_4$ has no systematic bias in the estimation of the parameters related to time trends, autocorrelations, or variances.

Note that $\beta_0$ expresses the baseline in local models, but $M_4$ also includes the term associated with elevation in the fixed effects, then the comparison of the posterior distribution of the intercept in local models and $\tilde{\beta}_0(\bs)$ is not of interest.

\subsection{Prediction at unobserved locations}

\begin{figure}[t]
	\centering
	\includegraphics[width=.4\textwidth]{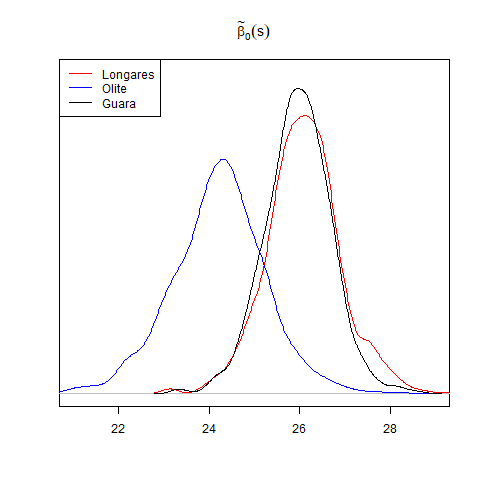}
	\includegraphics[width=.4\textwidth]{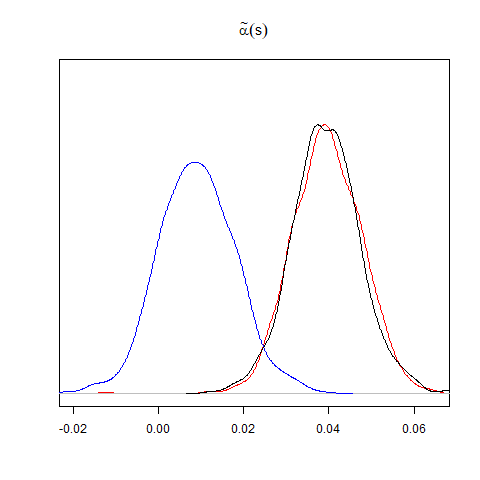}
	\includegraphics[width=.4\textwidth]{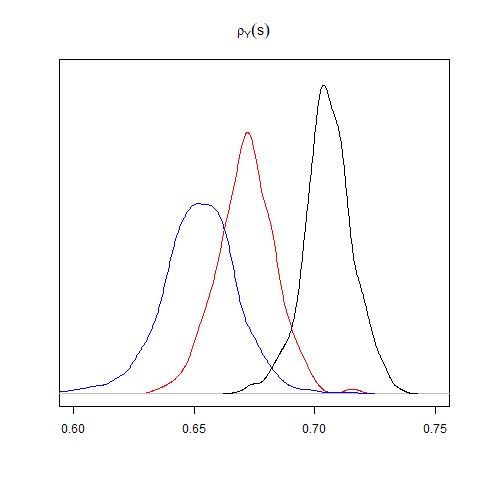}
	\includegraphics[width=.4\textwidth]{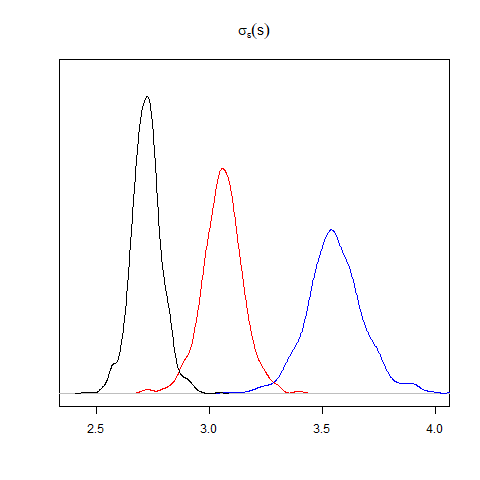}
	\caption{Posterior densities of the spatial random effects, $\tilde{\beta}_0(\bs),\tilde{\alpha}(\bs),\rho_Y(\bs),\sigma_\epsilon(\bs)$, at the unobserved locations Longares, Olite and Guara.}
	\label{fig:spatialGP}
\end{figure}

\begin{figure}[t]
	\centering
	\includegraphics[width=\textwidth]{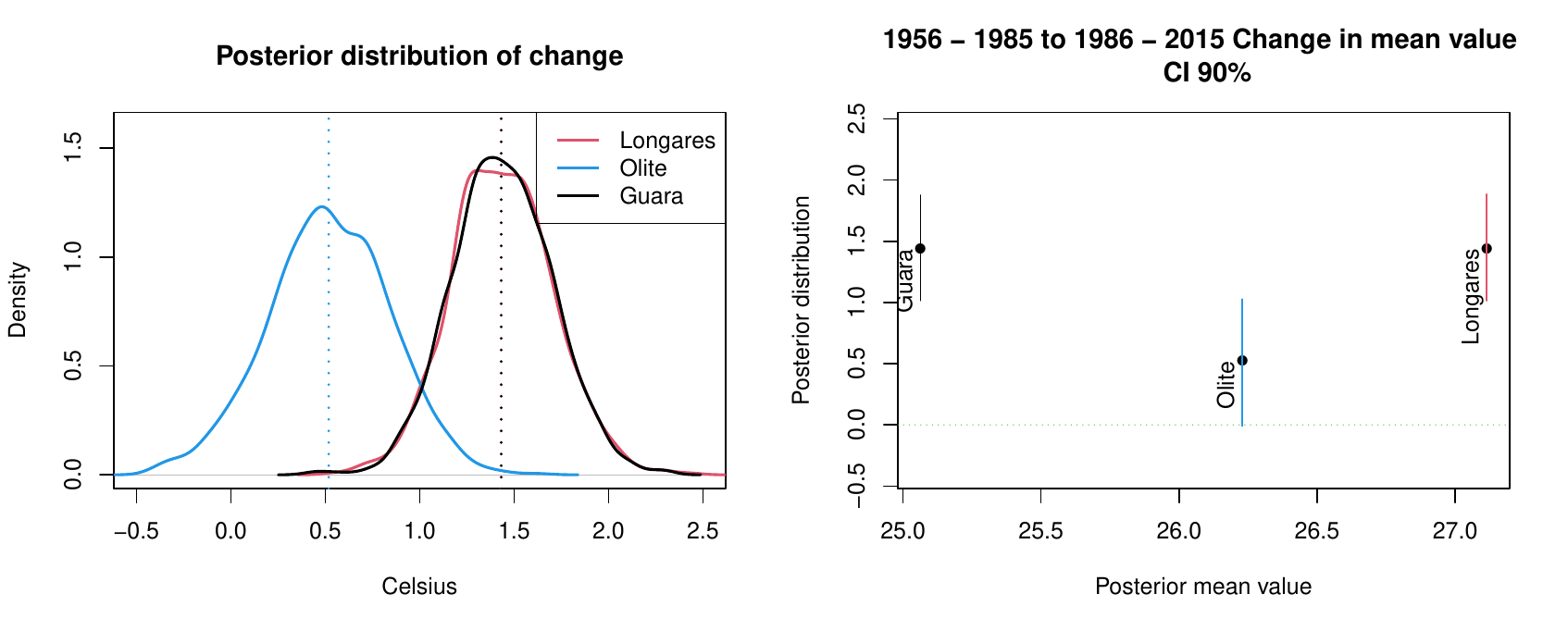}
	\caption{Left: Posterior densities of the difference between the mean value of the daily maximum temperature between both 30-year periods, 1956--1985 and 1986--2015, at the unobserved locations Longares, Olite and Guara. Right: Posterior mean and $90\%$ credible intervals of the previous difference against posterior mean value of the entire period.}
	\label{fig:LonOliGuaChange}
\end{figure}

Figure~\ref{fig:spatialGP} shows for each unobserved location (Longares, Olite and Guara) the posterior densities of the four spatial processes in $M_4$. Figure~\ref{fig:LonOliGuaChange} shows the posterior difference between average temperatures in both 30-year periods, 1956--1985 and 1986--2015.

\clearpage

\bibliographystyle{spbasic} 
\bibliography{ms}